\documentclass[aps,pra,reprint,10pt,a4paper]{revtex4-1}
\usepackage[utf8]{inputenc}
\usepackage[sc,osf]{mathpazo}
\usepackage{amsmath}
\usepackage[T1]{fontenc}
\usepackage{latexsym}
\usepackage{amssymb}
\usepackage[colorlinks=true,citecolor=blue,urlcolor=blue]{hyperref}
\usepackage{color}
\usepackage{graphics,epstopdf}
\usepackage{soul}
\usepackage[demo]{graphicx}
\usepackage{capt-of}
\usepackage{lipsum}
\usepackage{adjustbox}
\usepackage[normalem]{ulem}
\usepackage[table,xcdraw]{xcolor}
\usepackage{braket}
\usepackage{physics}
\usepackage{ragged2e}

\usepackage{comment}
\usepackage[font=small,labelfont=bf,justification=justified,singlelinecheck=false]{caption}

\begin{document}

\title{Information theoretic resource-breaking channels\\
%: The case for dense coding and teleportation
}

\author{Abhishek Muhuri$^{1,2}$, Ayan Patra$^1$, Rivu Gupta$^{1,3}$, Aditi Sen(De)$^1$}

\affiliation{$^1$ Harish-Chandra Research Institute,  A CI of Homi Bhabha National Institute, Chhatnag Road, Jhunsi, Prayagraj - 211019, India\\
$^2$ Center for Quantum Science and Technology, International Institute of Information Technology, Hyderabad - 500032, India \\
$^3$ Dipartimento di Fisica “Aldo Pontremoli,” Università degli Studi di Milano, I-$20133$ Milano, Italy}

\begin{abstract}
We propose the notion of \textit{process resource-breaking channels} that break the resource for a quantum information processing task. We examine the same using quantum dense coding and teleportation protocols. We prove that the sets $DBT$ (dense coding breaking) and $TBT$ (teleportation breaking) are convex and compact and identify classical-quantum channels as their extreme points. We prove group-covariance to be a sufficient condition for channels to be $DBT$ or $TBT$ when they can destroy the resource of maximally entangled states. \textcolor{black}{We present necessary and sufficient conditions for unital channels to be DBT for a single sender-receiver pair, while for multiple senders, the condition is sufficient.} The set of qubit $TBT$ channels is proved equivalent to qubit entanglement-breaking channels provided pre-processing is allowed.  We construct witness operators to identify non-$TBT$(non-$DBT$) maps.
\end{abstract}

\maketitle

\section{Introduction}
\label{sec:intro} 

Quantum information processing tasks are intended to demonstrate that quantum mechanical principles can significantly improve existing technologies. \cite{nielsen_2010, Preskill}. Additionally, the area has gained momentum since all proposed methods have been implemented in laboratories utilizing several physical substrates  \cite{Jones_NMR_Qcomputer_2000,  Maciej_coldatomreview_2007, Kok_linear_optical_Qcomputer_RMP_2007, Zukowski12,  Monroe_trapedion_RMP_2021}. The information flow, which may include the transmission of classical information or quantum states, occurs solely through classical or quantum channels in all of these architectures. Therefore, quantum channels \cite{nielsen_2010,Verstraete_2003,wilde_2013,Watrous_2018}, which are defined as completely positive and trace-preserving ($CPTP$) linear maps \cite{krauss_2013, Ruskai_LAA_2002}, that take density matrices (Hermitian, positive semidefinite operators with unit trace) from the input Hilbert space to other density matrices in the output Hilbert space, represent one of the fundamental building blocks of quantum information theory. 
Furthermore, the impact of quantum channels on states, which may or may not be connected to the environment, can, in principle, influence the protocols, and is therefore critical to analyse.  A plethora of studies have been conducted in this field, starting with the seminal work by dePillis \cite{dePillis_PJM_1967}, followed by the indispensable channel-state duality relation by Jamiolkowski \cite{Jamiolkowski_RMP_1972} and Choi \cite{Choi_LAA_1975} (for extension, see \cite{Paulsen_JMP_2013,Kye_JMP_2022,sohail_2023_arxiv}, and for geometric interpretation, see \cite{cortese_2002_arxiv,Petz_JMP_1996, Petz_AMH_2007, Petz_2008, Hiai_CMP_1991,Imai_STACS_1987}).
In recent years, some intriguing properties such as superadditivity \cite{Bennett_PRL_1997, DiVincenzo_PRA1998, Hastings_Nature_2009}, and superactivation \cite{Smith_Science_2008, Duan_arXiv_2009, Smith_NP_2011, Cubitt_IEEE_2011, Brandao_IEEE_2013} of quantum channels in terms of their ability to transfer information have been found in both finite \cite{Holevo_PPI_1973, Holevo_PPI_1979, Schumacher_PRA_1997, LLoyd_PRA_1997, Holevo_IEEE_1998, Bennett_IEEE_2002, Devetak_CMP_2005, Shirokov_CMP_2006, Shirokov_TVP_2008, Shirokov_PRA_2018, Shirokov_JMP_2019} and infinite dimensional systems \cite{Holevo_TPA_2006, Holevo_PIT_2008, Shirokov_PIT_2008, Holevo_DM_2010, Holevo_PIT_2010, Holevo_JMP_2011, Holevo_JMP_2016}.

Three classes of quantum channels, namely unital, group-covariant, and entanglement-breaking channels ($EBT$),  play significant roles in quantum information science and are also relevant to our work. For example, the set of unital channels, which can be completely characterized for qubits \cite{Tregub_IVUZM_1986, Landau_LAA_1993, Mendl_CMP_2009, Haagerup_AHP_2021, Girard_CMP_2022, Li_arXiv_2023}, is connected to operator quantum error correction \cite{Kribs_2005, Bacon_2006, Poulin_2005, Nielsen_2007} while for group covariant channels, which mimic the underlying symmetries of a system, the strong converse theorem for channel coding showing asymptotically vanishing probability for correct decoding was proven \cite{Konig_PRL_2009, Datta_QIP_2016} (see also \cite{Marvian_PRA_2014}, illustrating its role for the asymmetry properties of states). On the other hand, entanglement-breaking channels, which map any entangled state to separable ones \cite{Ruskai_RMP_2003, Horodecki_RMP_2003, Filipov_PRA_2012, Filipov_PRA_2013} and their properties including their classical capacities \cite{Shor_JMP_2002} have been extensively analyzed.

All quantum channels, explored in literature, have some intriguing qualities. For example, unital channels \cite{Li_arXiv_2023} play a vital role in state discrimination \cite{Bae_JPA_2015}, while group-covariant channels \cite{Memarzadeh_PRR_2022} are essential for studying relative entropy bounds on private communication \cite{Pirandola_Nature_2017}, for channel discrimination \cite{Jencova_JMP_2016}, and for the understanding of entanglement in spin systems \cite{Schliemann_PRA_2003}. $EBT$ channels \cite{Horodecki_RMP_2003}, non-locality breaking channels \cite{Pal_JPA_2015}, and coherence-breaking channels \cite{Luo_QIP_2022} on the other hand, destroy resources serving as fundamental ingredients in quantum information processes.  To begin with, we notice that for a particular job, a particular kind of quantum resource is required to outperform its equivalent classical strategy in terms of efficiency, thereby ensuring quantum advantage for that endeavor \cite{Bennett_PRL_1992, Bennett_PRL_1993, Murao_PRA_1999, Bennett_TCS_2014, Briegel_NP_2009, Giovannetti_NP_2011, Montanaro_npj_2016}. 
In this article, we introduce a distinct approach for characterizing quantum channels in relation to quantum information protocols.
Specifically, we analyze sets of channels that, while operating on states useful for a fixed protocol, convert them into states that are detrimental to that scheme -- we refer to these as ``process resource-breaking'' channels ($PBT$). The quantum protocols considered in this work are quantum dense coding (DC) \cite{Bennett_PRL_1992}, and quantum teleportation (QT) \cite{Bennett_PRL_1993}, and the corresponding resource breaking channels are referred to as ``dense coding resource-breaking'' ($DBT$) and ``teleportation resource-breaking'' ($TBT$) maps. 
It is crucial to emphasize that the sets of $PBT$s generally possess distinct qualities than those of $EBT$. Regardless of the choice of DC and QT in any arbitrary dimensions, we demonstrate that the set of such \(PBT\)s is convex and compact, and if $DBT$ and $TBT$ render all pure states useless, they also do the same for mixed states. Group-covariant channels that can eliminate the resource present in maximally entangled (ME) states for these schemes, can be universal $DBT$ and $TBT$ maps.
\textcolor{black}{Necessary and sufficient conditions for unital DBT channels are presented involving a single sender and receiver, while the condition is shown to be sufficient for multiple senders.} We further prove that if local pre-processing is permitted, qubit $TBT$ channels must be $EBT$. We also characterize the extreme points of the aforementioned channels and exhibit explicit construction of witness operators to identify the channels that cannot break the resources required for DC and QT.

Our paper is organized as follows. Sec. \ref{sec:PBT_definition} introduces the notion of {\it process resource-breaking channels}. In Sec. \ref{sec:properties_of_BT}, we define general features of channels that can serve as $DBT$ and $TBT$ and characterize their extreme points. We then analyze $DBT$ and $TBT$ channels independently in Secs. \ref{sec:DBT} and \ref{sec:TBT} respectively with special emphasis on their action on qubit systems. The penultimate discussion concerns the construction of witness operators for the proposed channels in Sec. \ref{sec:witness}. We end our manuscript with discussions and open questions in Sec. \ref{sec:conclu}.

\section{Process resource-breaking channels} 
\label{sec:PBT_definition}

We now propose the concept of channels that render states unsuitable for quantum information theoretic protocols.  For completeness, we refer the reader to Appendix \ref{app:pre}. for a detailed description of our notations and protocol definitions.
%We refer the reader to~\cite{Supp} for a detailed description of our notations. 
Consider a protocol, $\mathcal{P}$, with a performance quantifier, $\mathcal{Q}_\mathcal{P}$. Without quantum resources, $\mathcal{Q}_\mathcal{P} = \mathcal{Q}^{cl}_\mathcal{P}$, which we refer to as the classical bound, having no quantum advantage. If a state, $\rho$, can contribute such that $\mathcal{Q}_\mathcal{P}>\mathcal{Q}^{cl}_\mathcal{P}$, we deem it beneficial for $\mathcal{P}$ and denote the set of valuable states as $S_{\mathcal{P}}^{\rho_Q}$. Our aim is to characterize channels, $\Lambda$, which can transform such useful states into useless ones. We refer to the set of such channels as "{\bf process resource-breaking}" ($PBT$), which is mathematically equivalent to $(\Lambda\otimes\mathcal{I})\rho = \rho_{\Lambda} \notin S_{\mathcal{P}}^{\rho_Q}$. Here, $\mathcal{I}$ denotes the identity map, i.e., $\mathcal{I}(\rho) = \mathbb{I} \rho \mathbb{I}^\dagger$, with $\mathbb{I}$ being the identity operator.
%Note that instead of communication protocols, one can consider any quantum information processing task and can characterize channels that can lead to states that are non-advantageous for that particular task. 

It has been established that pre-processing can help to enhance the performance of protocols like dense coding and teleportation~\cite{Gupta_PRA_2021}. It prompts an immediate investigation into whether there exist another set of channels that, even in the presence of pre-processing operations, have the ability to destroy the associated resource. We call them "{\bf optimal process resource-breaking}" ($OPBT$)  channels. Mathematically,  $\Lambda_{p} (\Lambda\otimes\mathcal{I})\rho\notin S_{\mathcal{P}}^{\rho_Q}$, where $\Lambda_p$ refers to the pre-processing map. Note that $\Lambda \in OPBT \implies \Lambda \in PBT$. In our work, we refer to protocols without pre-processing as "{\bf standard}" protocols, and those aided by $\Lambda_p$ as "{\bf optimal}" ones.

Let us now focus on the setup of process resource-breaking channels in terms of dense coding, and quantum teleportation. The figure of merit for DC is the dense coding capacity, $C(\rho_{\mathcal{S}\mathcal{R}})=\max\{\log_2 d_{\mathcal{S}},~\log_2 d_{\mathcal{S}}+S(\rho_\mathcal{R})-S(\rho_{\mathcal{S}\mathcal{R}})\}$~\cite{Hiroshima_JPA_2001, Bruss_PRL_2004} of the state $\rho_{\mathcal{S}\mathcal{R}}$, where $\mathcal{S} = \mathcal{S}_1 \cdots \mathcal{S}_N$ denotes the collection of $N$ senders having dimensions $d_{\mathcal{S}} = d_{\mathcal{S}_1} \times \cdots \times d_{\mathcal{S}_N}$, $\mathcal{R}$ is the receiver system with dimension $d_{\mathcal{R}}$ and $S(\rho)=- \Tr(\rho \log_2 \rho)$ is the von Neumann entropy of $\rho$~\cite{nielsen_2010}. Similarly, the performance of optimal QT, is assessed via the maximum singlet fraction, $f_{\max}(\rho)=\max\{1/d,~\max_{L\in\text{LOCC}}\langle\phi^+|L(\rho)|\phi^+\rangle$\}~\cite{Horodecki_PRA_1999}, where $\rho$ is pre-shared between two parties having local dimension $d$ and $\ket{\phi^+}=\sqrt{1/d}\sum_{i=0}^{d-1}\ket{ii}$ with $\{\ket{i}\}_{i=0}^{d-1}$ being the computational basis in $\mathbb{C}^d$.
If $\rho$, acting on the Hilbert space $\mathbb{C}^d\otimes\mathbb{C}^d$, provides quantum advantage, $C(\rho)>\log_2 d_{\mathcal{S}}$ or $f_{\max}(\rho)>1/d$.

We refer to channels $\Lambda$ as "{\bf dense coding resource-breaking}" ($DBT$) channels when $C(\rho_{\mathcal{S}\mathcal{R},\Lambda})=C^{cl}\equiv\log_2d_{\mathcal{S}}~~\forall~~\rho_{\mathcal{S} \mathcal{R}}$ acting on $\mathbb{C}^{d_\mathcal{S}} \otimes \mathbb{C}^{d_\mathcal{R}}$ where $\rho_{\mathcal{S} \mathcal{R},\Lambda} = (\Lambda \otimes \mathcal{I}_{\mathcal{R}}) \rho_{\mathcal{S} \mathcal{R}}$. On the other hand, a channel $\Lambda$ is  "{\bf teleportation resource-breaking}" ($TBT$)  if $f_{\max}(\rho_{\mathcal{S} \mathcal{R},\Lambda}) = f^{cl} \equiv \frac{1}{d}~~\forall~~\rho_{\mathcal{S} \mathcal{R}}$ acting on $\mathbb{C}^d \otimes \mathbb{C}^d$. Since entanglement is a necessary resource for both protocols, $EBT$ channels are trivial examples of both $DBT$ and $TBT$. The goal of the present work is to identify $\Lambda \notin EBT$, but $\Lambda \in DBT(TBT)$. 

\textcolor{black}{\textbf{Note $\mathbf{1}$.} The dense coding protocol involves the use of two channels for its implementation. Consider a scenario involving a noisy DC protocol, where the receiver initially holds a two-qudit $(\mathbb{C}^d\otimes\mathbb{C}^d)$ entangled state $\rho_{\mathcal{R}' \mathcal{R}}$, which is subsequently transmitted to Alice via the channel $\mathcal{E}_1^{\mathcal{R}'\to \mathcal{S}}$. Alice then encodes her portion of the state using the operation $\Xi^{\mathcal{S} \to \mathcal{S}}: \rho_{\mathcal{S} \mathcal{R}} \to \sum_{i} q_i \Xi_i \otimes \mathbb{I} \rho_{\mathcal{S} \mathcal{R}} \Xi_i^\dagger \otimes \mathbb{I}$. The encoded state is sent back to the receiver through another channel $\mathcal{E}_2^{\mathcal{S} \to \mathcal{R}'}$. As a result of this process, the receiver ends up with an ensemble described by $\{q_i, \tilde{\rho}_{\mathcal{R}' \mathcal{R}}^i\}$, where $\tilde{\rho}_{\mathcal{R}' \mathcal{R}}^i=\mathcal{E}_2 \circ \Xi^i \circ \mathcal{E}_1\otimes \mathcal{I}(\rho_{\mathcal{R}' \mathcal{R}})$. The dense coding capacity, $C(\rho_{\mathcal{R}' \mathcal{R}})$, achievable in this protocol, can be determined by maximizing the Holevo quantity $\chi(\{q_i, \tilde{\rho}_{\mathcal{R}' \mathcal{R}}^i\})$ over the parameters $\{q_i, \Xi_i\}$. In our work, we consider the $DBT$ properties of $\mathcal{E}_1$ and $\mathcal{E}_2$ independently. Indeed, $\mathcal{E}_1 \in DBT \implies C(\rho_{\mathcal{R}' \mathcal{R}}) < \log_2 d_{\mathcal{S}}$, regardless of the nature of the channel $\mathcal{E}_2$. Thus, to analyse the $DBT$ characteristics of $\mathcal{E}_1$, we may assume $\mathcal{E}_2$ to be a noiseless channel.}

\section{Properties of communication resource-breaking channels}
\label{sec:properties_of_BT}

Herein we enlist ubiquitous features of \textbf{standard} dense coding and teleportation resource-breaking channels, independent of system dimensions.

%All the properties presented here hold for both $DBT$ and $TBT$, and we will collectively refer to these channels as .
\textbf{Theorem $\mathbf{1}$.} \textit{$DBT(TBT)$ channels satisfy the following properties: }

\begin{enumerate}
    \item Channels making pure states useless for standard $DC$ and $QT$, do the same for mixed states.
    \item Channels breaking the standard dense codeability or standard teleportability of $\ket{\phi^+}$ break the same for other ME states.
    \item  The set of $DBT(TBT)$ channels is convex and compact. \\
    $(a).$ The set $OTBT$ is convex and compact.
    %\hspace{10em}$(a).$ The channels $\Lambda \in OTBT$ form a convex and compact set.
    \item A channel breaking the dense codeability of a state before encoding, also does so if applied after the encoding process.    
    \item Extreme classical-quantum channels satisfying $\langle \psi_j|\psi_k\rangle\neq 0$ $\forall$ $j, k$ are extreme points of the set of $DBT(TBT)$ channels.
\end{enumerate}

\textit{Proof.} \textbf{Property $1$.} Any mixed state, $\rho$, can be decomposed into a convex combination of pure states as $\rho = \sum_i p_i \rho_i$ with $\sum_i p_i = 1$, where $\rho_i = \ket{\psi_i}\bra{\psi_i}$. The channel, being linear, acts on the mixed state as $\rho_\Lambda = \sum_i p_i \rho_{i, \Lambda}$.
%\begin{equation}
 %   (\Lambda \otimes \mathbb{I}) \rho = \sum_i p_i (\Lambda \otimes \mathbb{I}) \rho_i.
  %  \label{eq:mixed-broken}
%\end{equation}
Since we assume that the channel turns all pure states useless for {\bf standard} dense coding or teleportation, each state $\rho_{i, \Lambda}$ also becomes resource-less for those tasks. Note that the set of states that cannot offer any quantum advantage in the {\bf standard} dense coding or {\bf standard} teleportation protocol form a convex set \cite{Nirman_PRL_2011,Vempati_PRA_2021}. Therefore, a convex mixture of such states is not beneficial for $DC$ and $QT$. Hence the proof. $\hfill \blacksquare$

\textbf{Property $2$.} Let us first note that local unitary transformations can convert the state $\ket{\phi^+}$ to other ME states, as 
\begin{eqnarray}
    \nonumber \rho_{\text{ME}} && = (U\otimes V)\rho_{\phi^{+}}(U^{\dagger} \otimes V^{\dagger})\\
    && = (\mathbb{I}\otimes VU^{T})\rho_{\phi^{+}}(\mathbb{I}\otimes U^{T^{\dagger}}V^{\dagger}) \label{eq:ME_bell},
\end{eqnarray}
where, in the second step, we have used  the property $ (\mathcal{A}\otimes \mathbb{I}) |\phi^{+}\rangle = (\mathbb{I}\otimes \mathcal{A}^{T})|\phi^{+}\rangle$, unique to $\ket{\phi^+}$, for any operator $\mathcal{A}$.
The action of the channel, $\Lambda$, on $\rho_{\text{ME}}$ may be represented as
\begin{eqnarray}
    \nonumber \rho_{\text{ME},\Lambda}  = (\Lambda \otimes \mathcal{I}) \rho_{\text{ME}} = (\mathbb{I}\otimes VU^{T})\rho_{\phi^{+}, \Lambda}(\mathbb{I}\otimes U^{T^{\dagger}}V^{\dagger}).\\ \label{eq:noisy_ME}
\end{eqnarray}
Consequently, it follows that $C(\rho_{\text{ME},\Lambda})=C(\rho_{\phi^{+}, \Lambda})=\log_2d_S,$ where the equality is due to the fact that $\Lambda$ makes $\ket{\phi^+}$ useless for the dense coding protocol \textcolor{black}{and that the capacity is local unitarily invariant}.
\begin{comment}
In order to estimate the dense coding capacity, we consider the capacity expression corresponding to the noisy state in Eq. \eqref{eq:noisy_ME}.
\begin{eqnarray}
    && \nonumber \sum_j q_j S(\rho_{\text{ME},\Lambda}^j || \bar{\rho}_{\text{ME},\Lambda}) \\
    && = \sum_j q_j S(\rho_{\phi^+,\Lambda}^j||\bar{\rho}_{\phi^+,\Lambda}) \leq \log_2 d_{\mathcal{S}}.
    \label{eq:property2_dcc}
    \end{eqnarray}
    The last line follows from Eq. \eqref{eq:noisy_ME} upon using the monotonicity of the relative entropy function under local unitary operations and the inequality is due to the fact that $\Lambda$ makes $\ket{\phi^+}$ useless for the dense coding protocol. 
\end{comment}

   On the other hand, in case of the {\bf standard} teleportation protocol, the singlet fraction of the noisy ME state reads
   \begin{eqnarray}
       && \nonumber  \max_{U } \bra{\phi^+} U \rho_{\text{ME},\Lambda} U^\dagger \ket{\phi^+} \\
       && =  \max_{U} \bra{\phi^+} U \rho_{\phi^+,\Lambda} U^\dagger \ket{\phi^+} \leq 1/d_\mathcal{S},
       \label{eq:property2_tele}
   \end{eqnarray}
   where we have absorbed the local unitary transformations from Eq. \eqref{eq:noisy_ME} into the local unitary map $U$ in order to obtain the second equality. Eq. \eqref{eq:property2_tele} stems from the assumption that $\Lambda$ destroys the teleportation power of $\ket{\phi^+}$. Therefore, the proof follows. $\hfill \blacksquare$

\textbf{Property $3$.} We prove this statement in two parts.

First, we will show that {\bf {\bf standard}} $DC$ and $QT$ resource-breaking channels form a convex set. To do so, it is enough to demonstrate that the convex combination of any two arbitrary resource-breaking channels also represents a channel in $PBT_{D,T}$. Let $\Lambda_1, \Lambda_2 \in PBT_{D,T}$, whose convex combination reads $\Lambda = p_1 \Lambda_1 + p_2 \Lambda_2$ with $p_1 + p_2 = 1$.
Applying $(\Lambda \otimes \mathcal{I})$ on an arbitrary density operator $\rho$ gives
\begin{eqnarray}
    && \nonumber  \rho_{\Lambda} = p_1 (\Lambda_1 \otimes \mathcal{I})\rho + p_2(\Lambda_2 \otimes \mathcal{I})\rho \\
    && \implies  \rho_{\Lambda} = p_1 \rho_{\Lambda_1} + p_2 \rho_{\Lambda_2}
    \label{eq:convex_lambda_state}
\end{eqnarray}
Since $\Lambda_1$ and $\Lambda_2 \in PBT_{D,T}$, $\rho_{\Lambda_1}$ and $\rho_{\Lambda_2}$ cannot provide any quantum advantage in the considered protocols. Since states that cannot be used for the {\bf {\bf standard}} dense coding or teleportation tasks form a convex set, it implies that $\rho_\Lambda$ is a state that is useless as a resource in the {\bf {\bf standard}} dense coding or teleportation protocols. Hence the proof.

Let us now proceed to prove that the set $PBT_{D,T}$ is compact. 
%In this regard, we will employ a similar technique used in Ref. \cite{Nirman_PRA_2014}. 
Let $\Lambda_0$ be an arbitrary limit point of $PBT_{D,T}$
%. Such a limit point
which can always be identified since 
the given channels form a convex set, and there exists an infinite number of channels for different values of $p_1$ in Eq. \eqref{eq:convex_lambda_state}. Further, we construct a sequence $\{\Lambda_n \in B_n(\Lambda_0) \cap PBT_{D, T}\}$  of distinct $PBT_{D,T}$ channels around $\Lambda_0$, such that the channel $\Lambda_n$ resides in an open ball $B_n(\Lambda_0)$ of radius $1/n$ (which decreases with increasing $n$, implying $\lim_{n \to \infty} \Lambda_n \to \Lambda_0$ by construction) \cite{Nirman_PRA_2014}.
%\begin{eqnarray}
 %   && \nonumber \Lambda_1 \in B_1(\Lambda_0)\cap PBT_{D,T} \\
  %  && \nonumber \Lambda_2 \in B_{1/2}(\Lambda_0)\cap PBT_{D,T} \\
   % && \nonumber ...\\
    %&& \Lambda_n \in B_{1/n}(\Lambda_0)\cap PBT_{D,T}.
    %\label{eq:compact_seq}
%\end{eqnarray}
Let us now consider another sequence $\{(\mathcal{I}\otimes \Lambda_n)\rho = \chi_n\}$ for some arbitrary density operator $\rho$. Since $\Lambda_n \in PBT_{D,T}$, the state $\chi_n$ cannot be used to obtain quantum advantage in either the {\bf {\bf standard}} dense coding or the {\bf {\bf standard}} teleportation tasks. Hence, the sequence $\{\chi_n\}$ lies in the set of unsuitable states for {\bf {\bf standard}} dense coding or teleportation, having the limit point $\chi_0=(\mathcal{I}\otimes \Lambda_0)\rho$, since
%\begin{eqnarray}
% && \nonumber 
 \(\lim_{n \to \infty}  (\mathcal{I}\otimes \Lambda_n)\rho \to (\mathcal{I}\otimes \Lambda_0)\rho \)
    \(\implies \lim_{n \to \infty} \chi_n \to \chi_0\).
  %  \label{eq:limit_compact_seq}
%\end{eqnarray}
 As the states which cannot support {\bf {\bf standard}} dense coding or teleportation form a closed set \cite{Nirman_PRL_2011, Vempati_PRA_2021}, they contain the limit point $\chi_0$. This, in turn, implies that the limit of the arbitrary sequence of channels in $PBT_{D,T}$ is also a resource-breaking channel.
 %, through Eq. \eqref{eq:limit_compact_seq}. 
 Therefore, $\Lambda_0 \in PBT_{D,T}$ and since we have considered $\Lambda_0$ to be an arbitrary limit point, we can conclude that the set of {\bf {\bf standard}} $DC$ and $QT$ resource-breaking channels contains all its limit points and is hence closed. Note that $PBT_{D,T}\subseteq CPTP$ with the latter being a compact set. Using the fact that a closed subset of a compact set is also compact, and hence the proof. $\hfill \blacksquare$

\textcolor{black}{\textbf{Property $3(a)$.} \textbf{Convexity.} The {\it optimal} teleportation protocol is characterized by optimizing the singlet fraction over all LOCC protocols, $L$, i.e.,
\begin{equation}
    f_{\max}(\rho_{\mathcal{S} \mathcal{R}}) = \max_{L \in \text{LOCC}} \bra{\phi^+} L(\rho_{\mathcal{S} \mathcal{R}}) \ket{\phi^+}.
\end{equation}
Let us consider two states, $\rho_1$ and $\rho_2$ acting on $\mathbb{C}^d \otimes \mathbb{C}^d$, such that $f_{\max} \leq 1/d$ for both. The optimal singlet fraction corresponding to $\rho = p_1 \rho_1 + p_2 \rho_2$ (with $p_1 + p_2 = 1$) reads
\begin{eqnarray}
    f_{\max}(\rho) =  \bra{\phi^+} L_{\rho}(p_1 \rho_1 + p_2 \rho_2) \ket{\phi^+},
\end{eqnarray}
where $L_{\rho}$ denotes the optimal LOCC protocol that attains the maximum singlet fraction for $\rho$. Hence, we obtain
\begin{eqnarray}
\label{eq:singlet_fraction_convexity}
    \nonumber  f_{\max}(\rho) && = p_1 \bra{\phi^+} L_{\rho}(\rho_1) \ket{\phi^+}  + p_2 \bra{\phi^+} L_{\rho}(\rho_2) \ket{\phi^+} \\
    &&\leq p_1f_{\max}(\rho_1) + p_2f_{\max}(\rho_2)\nonumber\\
  &&  \implies f_{\max}(\rho) \leq \frac{1}{d}.
\end{eqnarray}
\textcolor{black}{It is important to highlight that, because the set of LOCC operators is not closed, two scenarios can arise: either an optimal LOCC operator, denoted as $L_{\rho}$, corresponding to $\rho$, exists, or it does not. In the first scenario, we can consider an infinite sequence of LOCC operators for $\rho$, denoted by $\{L_{\rho}^i\}$, such that $\bra{\phi^+}L_{\rho}^i(\rho)\ket{\phi^+}\leq \bra{\phi^+}L_{\rho}^{i+1}(\rho)\ket{\phi^+}$ and $\lim_{i\to\infty} L_{\rho}^i = L_{\rho}$. In the second scenario, an arbitrarily large but finite sequence of LOCC operators exists, satisfying the same monotonic condition. It can be easily verified that Eq. (\ref{eq:singlet_fraction_convexity}) holds for all elements of that sequence (with $L_{\rho}$ replaced by $L_{\rho}^i$ in Eq. (\ref{eq:singlet_fraction_convexity})), thereby preserving the convexity property.} Thus, $\rho$ that fails to attain quantum advantage in the optimal teleportation protocol form a convex set. Therefore, channels that can destroy the resource in the optimal teleportation protocol also form a convex set.\\\\
\textcolor{black}{\textbf{Compactness.} Let us suppose that for any two density matrices $\rho_1$ and $\rho_2$, $f_{\max}(\rho_i)$ is achieved at $L_{\rho_i}$. Therefore, we have $f_{\max}(\rho_1)-f_{\max}(\rho_2)=\bra{\phi^+} L_{\rho_1}(\rho_1) \ket{\phi^+}-\bra{\phi^+} L_{\rho_2}(\rho_2) \ket{\phi^+}\leq\bra{\phi^+} L_{\rho_1}(\rho_1) \ket{\phi^+}-\bra{\phi^+} L_{\rho_1}(\rho_2) \ket{\phi^+}=\bra{\phi^+} L_{\rho_1}(\rho_1-\rho_2) \ket{\phi^+}$. Since the set of $CPTP$ operators in finite dimension is compact, it is bounded. Hence the set of LOCC operators, being a subset of $CPTP$, is also bounded because every subset of a bounded set is also bounded. Although, it should be noted that the set of LOCC operators is not compact since it is not closed. As a consequence, we obtain $|f_{\max}(\rho_1)-f_{\max}(\rho_2)|\leq ||\ket{\phi^+}||^2||L_{\rho_1}||||\rho_1-\rho_2||=C||\rho_1-\rho_2||$, where $||(.)||$ is any operator norm and $C$ being an arbitrary constant. This implies that $f_{\max}(\rho)$ is a continuous function of $\rho$. Moreover $f_{\max}(\rho)\in\left[\frac{1}{d^2},\frac{1}{d}\right]$ for the states which are not useful for optimal teleportation. This concludes that the set of states which cannot be used to obtain quantum advantage in optimal teleportation protocol is closed. Therefore, channels that can destroy the resource in optimal teleportation also form a compact set. Hence the proof.} $\hfill \blacksquare$}

\textbf{Property $4$.} When the channel through which the encoded subsystems are sent is noisy, the dense coding capacity takes the form as \cite{Shadman_NJP_2010}
    \begin{eqnarray}
        &&  \nonumber   \tilde{C} = \max\Big[\log_2d_{\mathcal{S}_1} \cdots d_{\mathcal{S}_{N - 1}}, \\ 
      && \nonumber  \log_2d_{\mathcal{S}_1} \cdots d_{\mathcal{S}_{N - 1}} + S(\rho_{\mathcal{R}}) - S((\tilde{\Lambda}_{\mathcal{S}_1 \cdots \mathcal{S}_{N-1}} \otimes \mathcal{I}_{\mathcal{R}}) \times\\
      && \nonumber (U_{\mathcal{S}_1} \otimes \cdots \otimes U_{\mathcal{S}_{N-1}}\otimes \mathbb{I}_{d_{\mathcal{R}}}\rho_{\mathcal{S}_1 \cdots \mathcal{S}_{N - 1} \mathcal{R}} \times \\
      &&  U^\dagger_{\mathcal{S}_{1}} \otimes \cdots \otimes U^\dagger_{\mathcal{S}_{N-1}} \otimes \mathbb{I}_{d_{\mathcal{R}}}))\Big],
      \label{eq:noisy_dcc}
    \end{eqnarray}
    with $U_{\mathcal{S}_i}$ being unitaries applied by the sender $\mathcal{S}_i$ to minimize the entropy of the state upon the action of the noisy channel $\tilde{\Lambda}$. Note that $U_{\mathcal{S}_1} \otimes \cdots \otimes U_{\mathcal{S}_{N-1}} \otimes \mathbb{I}_{d_{\mathcal{R}}}\rho_{\mathcal{S}_1 \cdots \mathcal{S}_{N - 1} \mathcal{R}} U^\dagger_{\mathcal{S}_1} \otimes \cdots \otimes U^\dagger_{\mathcal{S}_{N-1}} \otimes \mathbb{I}_{d_{\mathcal{R}}} $ which denotes another valid state in $\mathcal{C}^{d_{\mathcal{S}}} \otimes\mathcal{C}^{d_{\mathcal{R}}}$. Let us denote it as $\hat{\rho}$. Therefore, the action of the noisy channel on the encoded qudit of $\rho_{\mathcal{S}_1 \cdots \mathcal{S}_{N - 1} \mathcal{R}}$ is mathematically equivalent to that on a unitarily rotated state $\hat{\rho}$, prior to encoding (since the noise acts only on the sender subsystem, $\rho_\mathcal{R}$ retains the same form, regardless of whether the noise acts before or after encoding). Since $\Lambda$ can break the {\bf {\bf standard}} dense coding capacity of all states when it acts before the implementation of the protocol, the proof follows. $~~~~~~~~~~~~~~~~~~~~~~~~~~~~~~~~~~~~~~~~~~~~~~~~~~~~~~~~~~~~~~~~~~~~~~~~~~~~~~~~~~~~~~~~~~\blacksquare$

\textbf{\textbf{Property $5$.}} 
Recall that the sets $PBT_{D,T}$, $EBT$, and $CPTP$ are all convex and compact. Moreover, $EBT \subseteq PBT_{D,T} \subseteq CPTP$, since $EBT$ channels are trivial examples of $PBT_{D,T}$ channels. The extreme $CQ$ channels are extreme points of $EBT$ \cite{Horodecki_RMP_2003} and, therefore, they also belong to the set of channels containing $PBT_{D,T}$. Again, extreme $CQ$ channels having the property, $\langle\psi_i|\psi_j\rangle\neq0~~\forall~~i,j$ are extreme points of $CPTP$ \cite{Horodecki_RMP_2003}. As a result, they will also be extreme points of $PBT_{D,T}$ because they belong to both $PBT_{D,T}$ and $CPTP$, with the hierarchy being $PBT_{D,T} \subseteq  CPTP$. Hence the proof.$\hfill \blacksquare$

Properties $1, 2$, and $3$ also hold for \textbf{"optimal teleportation resource-breaking"} ($OTBT$) channels, while it may not do so for \textbf{"optimal dense coding resource-breaking"} ($ODBT$) channels. Due to the second property, we will hereafter imply the action of $\Lambda$ on $\ket{\phi^+}$ whenever we consider the action of $DBT(TBT)$ channels on ME states. First, let us concentrate on resource-breaking for the dense coding protocol.

\section{Characterization of dense coding resource-breaking channels}
\label{sec:DBT} 

We now concentrate on channels that make dense codeable states useless for the {\bf standard} protocol. In particular, we provide a sufficient condition on channels that can destroy resource states in arbitrary dimensions. %We further consider the specific case of qubit systems and analyze sufficient properties of unital channels to be $DBT$.
Note that we are concerned with the functional form of the DC capacity, $C(\rho_{\mathcal{S}\mathcal{R}})=\log_2 d_{\mathcal{S}}-S_{\mathcal{S}|\mathcal{R}}(\rho_{\mathcal{S}\mathcal{R}})$, rather than maximizing over the classical threshold. Here, $S_{\mathcal{S}|\mathcal{R}}(\rho_{\mathcal{SR}})=S(\rho_{\mathcal{SR}})-S(\rho_\mathcal{R})$ is the conditional von Neumann entropy.

%\textbf{Theorem 5.} \textit{If the action of a channel makes the maximally entangled state non-dense codeable, the group covariance property is a sufficient condition for it to break the dense codeability of an arbitrary pure state.}\\

\textbf{Theorem $2$.} \textit{ Sufficient conditions on a channel, $\Lambda$, to be $DBT$ include\\
(a) group covariance with respect to $SU(d_{\mathcal{S}})$, and\\
(b) ability to make $|\phi^+\rangle$ non-dense codeable,\\
with $SU(d_{\mathcal{S}})$ being the group of special unitary operators in sender's dimension $d_{\mathcal{S}}$.}

Before proceeding further, we refer the reader to Appendix.~\ref{app:pre} for formal definitions of group-covariant, A-unital and conditional von Neumann entropy non decreasing maps, which we shall use in our proof. 

\textit{Proof.} According to Nielsen's protocol~\cite{Nielsen_PRL_1999}, $|\phi^+\rangle$ can be transformed into any general two-qudit pure state $\ket{\psi} \in \mathbb{C}^{d_{\mathcal{S}}} \otimes \mathbb{C}^{d_{\mathcal{R}}}$ $(\text{with }\rho_{\psi}=\ketbra{\psi}{\psi})$ as 
\begin{eqnarray}
  \nonumber \rho_{\psi} && = \sum_{\alpha} (\prod_{r=1}^{\lfloor \frac{d}{2}\rfloor} U_{\alpha_{1+\lfloor \frac{d}{2}\rfloor-r}}\otimes V_{\alpha_{1+\lfloor \frac{d}{2}\rfloor-r}}M_{\alpha_{1+\lfloor \frac{d}{2}\rfloor-r}})\rho_{\phi^+}\\
  \nonumber && \times (\prod_{r=1}^{\lfloor \frac{d}{2}\rfloor} U_{\alpha_r}^{\dagger}\otimes M_{\alpha_r}^{\dagger}V_{\alpha_r}^{\dagger})\\
   && =\sum_{\alpha} (U_{\alpha}\otimes \Tilde{M}_{\alpha})\rho_{\phi^+}(U_{\alpha}^{\dagger}\otimes \Tilde{M}_{\alpha}^{\dagger}),
   \label{eq:state-transf}
\end{eqnarray}
where $M_{\alpha_k}$ are POVMs, and  $U_{\alpha_k}$, $V_{\alpha_k}$ are local unitary operations~\cite{Torun_PLA_2015}, with $\alpha\equiv\{\alpha_1,\alpha_2,...,\alpha_{\lfloor \frac{d}{2}\rfloor}\}$. 
%Here, for brevity, we define $U_{\alpha}=\prod_{r=1}^{\lfloor \frac{d}{2}\rfloor} U_{\alpha_{1+\lfloor \frac{d}{2}\rfloor-r}}$ and $\Tilde{M}_{\alpha}=\prod_{r=1}^{\lfloor \frac{d}{2}\rfloor}V_{\alpha_{1+\lfloor \frac{d}{2}\rfloor-r}}M_{\alpha_{1+\lfloor \frac{d}{2}\rfloor-r}}$.\\
%\begin{eqnarray}
    %\rho_{\psi}&&= \sum_{\alpha} (U_{\alpha}\otimes V_{\alpha})(\mathbb{I}\otimes M_{\alpha})\rho_{\phi^+}(\mathbb{I}\otimes M_{\alpha}^{\dagger})(U_{\alpha}^{\dagger}\otimes V_{\alpha}^{\dagger})\\
    %&&= \sum_{\alpha} (U_{\alpha}\otimes V_{\alpha}M_{\alpha})\rho_{\phi^+}(U_{\alpha}^{\dagger}\otimes M_{\alpha}^{\dagger}V_{\alpha}^{\dagger})\\
%\end{eqnarray}
Further, $(\Lambda\otimes\mathcal{I})\rho_{\psi}=\rho_{\psi, \Lambda} = \sum_{\alpha} (\Lambda \otimes \mathcal{I}) (U_{\alpha}\otimes \Tilde{M}_{\alpha})\rho_{\phi^+}(U_{\alpha}^{\dagger}\otimes \Tilde{M}_{\alpha}^{\dagger}).$
%\begin{equation}
  %  \rho_{\psi, \Lambda} = \sum_{\alpha} (\Lambda \otimes \mathbb{I}) (U_{\alpha}\otimes \Tilde{M}_{\alpha})\rho_{\phi^+}(U_{\alpha}^{\dagger}\otimes \Tilde{M}_{\alpha}^{\dagger}).
   % \label{eq:noisy_arb_pure}
%\end{equation}
If $\Lambda$ is group-covariant, $(\Lambda \otimes \mathcal{I}) (U_{\alpha}\otimes \Tilde{M}_{\alpha})\rho_{\phi^+}(U_{\alpha}^{\dagger}\otimes \Tilde{M}_{\alpha}^{\dagger}) = (\tilde{U}_{\alpha}\otimes \Tilde{M}_{\alpha})(\Lambda \otimes \mathcal{I}) \rho_{\phi^+}(\tilde{U}_{\alpha}^{\dagger}\otimes \Tilde{M}_{\alpha}^{\dagger})$, where $U_\alpha, \tilde{U}_\alpha$ are $d$-dimensional representations of the compact Lie group, $SU(d_\mathcal{S})$. As a result, $\rho_{\psi,\Lambda}= \sum_{\alpha} (\Tilde{U}_{\alpha}\otimes \Tilde{M}_{\alpha})\rho_{\phi^+,\Lambda}(\Tilde{U}_{\alpha}^{\dagger}\otimes \Tilde{M}_{\alpha}^{\dagger}) = \Lambda'(\rho_{\phi^+,\Lambda})$,
%We apply a group covariant channel now on the 1st party of $\rho_{\alpha}$,
%\begin{eqnarray}
   %\nonumber \rho_{\psi,\Lambda}&&= \sum_{\alpha} (\Tilde{U}_{\alpha}\otimes \Tilde{M}_{\alpha})\rho_{\phi^+,\Lambda}(\Tilde{U}_{\alpha}^{\dagger}\otimes \Tilde{M}_{\alpha}^{\dagger})\\
    %&&= \Lambda'(\rho_{\phi^+,\Lambda}),
    %\label{eq:noisy_arb_pure_GC}
%\end{eqnarray}
where $\Lambda'\in LOCC$ is an A-unital map $(UNI(A|B))$ i.e., $(\Lambda'\otimes\mathcal{I})(\mathbb{I}_A\otimes\rho_B)=(\mathbb{I}_A\otimes\rho_B')~\forall~\rho_B$~\cite{Vempati_Quantum_2022}. The set of A-unital maps is equivalent to that of conditional von Neumann entropy non decreasing maps defined as $NCVE(A|B)=\{\Lambda:S_{A|B}(\Lambda(\chi_{AB})) \geq S_{A|B}(\chi_{AB})~\forall~\chi_{AB}\}$~\cite{Vempati_Quantum_2022}. So, $\Lambda'\in NCVE(A|B)$, and
%whose action on a state, say $\chi$, is given by $\Lambda'(\chi) = \sum_{\alpha} (\Tilde{U}_{\alpha}\otimes \Tilde{M}_{\alpha})\chi(\Tilde{U}_{\alpha}^{\dagger}\otimes \Tilde{M}_{\alpha}^{\dagger})$. 
%Evidently,  $\Lambda' \in UNI(A|B) \implies \Lambda' \in NCVE(A|B)$~\cite{Vempati_Quantum_2022}, i.e., for any $\chi_{\mathcal{S} \mathcal{R}}$, we have $S_{\mathcal{S}|\mathcal{R}}(\Lambda'(\chi_{\mathcal{S} \mathcal{R}})) \geq S_{\mathcal{S}|\mathcal{R}}(\chi_{\mathcal{S} \mathcal{R}})$. 
the dense coding capacity reads
\begin{eqnarray}
     \nonumber && C(\rho_{\psi, \Lambda}) = \log_2 d_{\mathcal{S}} - S_{\mathcal{S}|\mathcal{R}}(\rho_{\psi, \Lambda}) \\
     && \leq \log_2 d_{\mathcal{S}} - S_{\mathcal{S}|\mathcal{R}}(\rho_{\phi^+, \Lambda}) = C(\rho_{\phi^+, \Lambda}).
     \label{eq:g-cov-proof}
\end{eqnarray}
Therefore, if $\Lambda$ makes $\ket{\phi^+}$ non-dense codeable, its group covariant nature guarantees the same for all other pure states. $\hfill \blacksquare$

An example of dense coding resource-breaking for a class of group covariant channels is provided in Appendix.~\ref{app:gc-dcc-break}.

\textcolor{black}{\textbf{Remark $\mathbf{1}$.} The entanglement-assisted classical capacity of a channel, $\Lambda$, represents the highest rate at which classical information can be transmitted, with minimal error, when the sender and receiver utilize the channel along with an unlimited amount of shared entanglement. In Ref. \cite{Bennett_IEEE_2002}, it was shown that this capacity is given by $C_E(\Lambda)=\max_{\rho}I(A:B)_{\sigma}$, where the quantum mutual information is evaluated on the state $\sigma=\mathcal{I}\otimes\Lambda(\rho)$. Theorem $2$ suggests that this maximization may be attained at the maximally entangled state for the $SU(d_\mathcal{S})$ group covariant channels.}

\textcolor{black}{\textbf{Note $2$.} In optimal dense coding, if the pre-processing operation $\Lambda_p \in UNI(A|B)$, $\Lambda_p (\rho_{\psi, \Lambda}) = \Lambda_p(\Lambda' (\rho_{\phi^+, \Lambda}))$ with $\Lambda_p \circ \Lambda' \in UNI(A|B) \implies \Lambda\in ODBT$. An example of such pre-processing channel is $\Lambda_p  = \Lambda_{\mathcal{S}} \otimes \Lambda_{\mathcal{R}}$, with $\Lambda_{\mathcal{S}}$ being a unital map.}

\subsection{Qubit unital $DBT$ channels}
\label{subsec:qubit-unital-DBT}

We now focus particularly on $DBT$ qubit unital channels. Up to unitary pre-and post-processing, the action of any canonical unital qubit channel is $\Lambda_{UNI} (\xi)=\sum_{i=1}^4p_i\Lambda_i(\xi)$~\cite{Li_arXiv_2023} where $0\leq p_i\leq1,~~\sum_ip_i=1$ and $\nonumber \Lambda_1:\xi \to \frac{1}{2} \Tr(\xi) \mathbb{I}_2,~ \Lambda_2:\xi \to \xi,~ \Lambda_3:\xi \to \frac{1}{2} (\xi+\sigma_x \xi \sigma_x),~ \Lambda_4:\xi \to \frac{1}{2} (\sigma_x \xi \sigma_x + \sigma_y \xi \sigma_y).$
%\begin{eqnarray}
 %  \nonumber \Lambda_1:\chi \to \frac{1}{2} \Tr(\chi) \mathbb{I}_2, && \Lambda_2:\chi \to \chi,\\
 %   \Lambda_3:\chi \to \frac{1}{2} (\chi+\sigma_x \chi \sigma_x), && \Lambda_4:\chi \to \frac{1}{2} (\sigma_x \chi \sigma_x + \sigma_y \chi \sigma_y).\nonumber\\
 %   \label{eq:unital_channel_decomposition}
%\end{eqnarray}
Therefore, any qubit unital channel reads $\tilde{\Lambda}_{UNI}=U_2\circ\Lambda_{UNI}\circ U_1$. Since unitaries can never be $DBT$, unital channels become $DBT$ if and only if their canonical part is so, making it equivalent to work only with the canonical part in our context.

%If a unital channel acts on the sender's part of $\rho_{\mathcal{S} \mathcal{R}}$, the resulting noisy state is
%\begin{eqnarray}
%&& \nonumber \rho_{\mathcal{S} \mathcal{R}, \Lambda} =p_1\frac{\mathbb{I}_2}{2} \otimes \rho_{\mathcal{R}}+p_2 \rho_{\mathcal{S} \mathcal{R}}+ \frac{p_3}{2}(\rho_{\mathcal{S} \mathcal{R}}+\\
%&&\sigma_x\otimes\mathbb{I}~\rho_{\mathcal{S} \mathcal{R}}~\sigma_x\otimes\mathbb{I})+ \frac{p_4}{2}(\sigma_x\otimes\mathbb{I}~\rho_{\mathcal{S} \mathcal{R}}~\sigma_x\otimes\mathbb{I}+\nonumber\\&&\sigma_y\otimes\mathbb{I}~\rho_{\mathcal{S} \mathcal{R}}~\sigma_y\otimes\mathbb{I}),
%\label{eq:unital-qubit-ch}
%\end{eqnarray}
%where $\rho_{\mathcal{R}} = \Tr_\mathcal{S} (\rho_{\mathcal{S} \mathcal{R}})$. For brevity, we refer to $\sigma_i\otimes\mathbb{I}$ as $\sigma_i^A$.

\textcolor{black}{
 Let us recall the dense coding situation, involving the channels $\mathcal{E}_1$ (during state distribution), $\Xi : \rho \to \sum_i q_i \mathcal{W}_i \rho \mathcal{W}_i^\dagger$ (encoding with d-dimensional Pauli operators~\cite{Hiroshima_JPA_2001}), and $\mathcal{E}_2$ (to transmit the encoded state), as described at the end of Sec.~\ref{sec:PBT_definition}. As a result of this process, the receiver ends up with an ensemble described by $\{q_i, \tilde{\rho}_{\mathcal{R}' \mathcal{R}}^i\}$, where $\tilde{\rho}_{\mathcal{R}' \mathcal{R}}^i=\mathcal{E}_2 \circ \Xi_i \circ\mathcal{E}_1\otimes \mathcal{I}(\rho_{\mathcal{R}' \mathcal{R}})$. The maximum dense coding capacity of the channel, $C_D(\mathcal{E}_{1,2})$, achievable in this protocol can be determined by maximizing the Holevo quantity $\chi(\{q_i, \tilde{\rho}_{\mathcal{R}' \mathcal{R}}^i\})$ over the parameters $\{q_i, \rho_{\mathcal{R}' \mathcal{R}}\}$. When both channels $\mathcal{E}_1$ and $\mathcal{E}_2$ are \textit{Pauli covariant}, the capacity simplifies to $C_D(\mathcal{E}_{1,2}) = \max_{\{q_i\}}\chi(\{q_i, \mathcal{E}_2\circ\Xi_i\circ\mathcal{E}_1\otimes \mathcal{I}(\rho_{\phi^+})\})$ (see \cite{Laurenza_PRR_2020}). Moreover, if $\mathcal{E}_1$ and $\mathcal{E}_2$ are specifically \textit{Pauli channels}, defined by $\mathcal{E}_k(\rho)=\sum_i p_i^k\sigma_i\rho\sigma_i$, where $p_i^k \geq 0$ and $\sum_i p_i^k=1$ for $k\in\{1,2\}$, the maximization over $\{q_i\}$ is achieved at $q_i=\frac{1}{d^2}$. Therefore, for Pauli channels, the capacity is given by $C_D(\mathcal{E}_{1,2})=2\log_2d - S(\mathcal{E}_2\circ\mathcal{E}_1\otimes \mathcal{I}(\rho_{\phi^+}))=C(\rho_{\phi^+,\mathcal{E}_2\circ\mathcal{E}_1})$.}

\textcolor{black}{
\textbf{Theorem 3.} \textit{A unital qubit channel acting on one half of any two-qubit pure state is capable of making it non-dense codeable if and only if it can break the dense coding capacity of any ME state.}}

\textcolor{black}{
\textit{Proof.} As already discussed, any unital qubit channel is unitarily equivalent to some canonical unital channel. A slight manipulation reveals that this canonical channel is indeed a qubit Pauli channel, expressed as $\Lambda_P(\rho)=\sum_i p_i\sigma_i\rho\sigma_i$, where $\sigma_i\in\{\mathbb{I},\sigma_x, \sigma_y, \sigma_z\}$ and $\sum_i p_i=1$ with $p_i \geq 0$. This suggests that we can write an arbitrary qubit unital channel as $\tilde{\Lambda}_{UNI}=U_{2}\circ\Lambda_P\circ U_{1}$. Using the fact that the conditional entropy remains invariant under local unitary operation, we obtain $C(\rho_{\phi^+,\tilde{\Lambda}_{UNI}})=C(U_2\otimes U_1^T (\rho_{\phi^+,\Lambda_{P}}))=C(\rho_{\phi^+,\Lambda_{P}})$. Thus a unital qubit channel breaking the dense codeability of an ME state implies that the corresponding canonical Pauli channel does so. Now we use the result of Ref. \cite{Laurenza_PRR_2020} in the following way. Recall, that to independently probe the $DBT$ properties of a channel which acts before encoding, we can consider the second channel to be noiseless, i.e., $\mathcal{E}_2 = \mathcal{I}$, as discussed at the end of Sec.~\ref{sec:PBT_definition}. Consequently, we can express  $\max_{\{q_i,\rho_{\mathcal{R}' \mathcal{R}}\}}\chi(\{q_i,\Xi_i\circ\mathcal{E}_1\otimes\mathcal{I}\rho_{\mathcal{R}' \mathcal{R}}\})=\max_{\{\rho_{\mathcal{R}' \mathcal{R}}\}}C(\mathcal{E}_1\otimes\mathcal{I}(\rho_{\mathcal{R}' \mathcal{R}}))$, where only the maximization over the input states needs to be considered, since $\mathcal{E}_2 = \mathcal{I}$ means that Pauli encoding with $q_i = 1/4~\forall~i$ is optimal~\cite{Hiroshima_JPA_2001}, which leads to the dense coding capacity.  By substituting $\mathcal{E}_1$ with $\Lambda_P$, we obtain $C_D(\Lambda_P)=\max_{\{\rho_{\mathcal{R}' \mathcal{R}}\}}C(\Lambda_P\otimes\mathcal{I}(\rho_{\mathcal{R}' \mathcal{R}}))=C(\rho_{\phi^+,\Lambda_P})$. Therefore, $C(\rho_{\phi^+,\Lambda_P})\leq 1 \implies C(\rho_{\psi,\Lambda_P})\leq 1 ~ \forall~ \rho_\psi \in \mathbb{C}^2\otimes\mathbb{C}^2$, which further implies $\Lambda_P\in DBT$. Consequently, if a unital qubit channel acting on one half of a maximally entangled state breaks its dense coding capability, it transforms any state into one that is ineffective for dense coding, thereby becoming a $DBT$ channel. The converse is also trivially true by the definition of $DBT$ channels. This completes the proof.  $\hfill \blacksquare$}

\textcolor{black}{
\textbf{Remark $2$:} The entanglement-assisted classical capacity for a Pauli channel, $\Lambda_P^d$, acting on $\mathbb{C}^d$, is given by $2\log_2d-S(\Lambda_P^d\otimes\mathcal{I}(\rho_{\phi^+}))$~\cite{Bennett_IEEE_2002,Bennett_PRL_1999,Laurenza_PRR_2020}. Theorem $3$ allows us to extend this result to any unital qubit channel $\tilde{\Lambda}_{UNI}$, where the entanglement-assisted classical capacity is similarly expressed as $2\log_2 2-S(\tilde{\Lambda}_{UNI}\otimes\mathcal{I}(\rho_{\phi^+}))$.}

Let us turn to multi-qubit DC with $N-1$ senders sharing an $N$-qubit system with a single receiver, where any $m (< N-1)$ different senders are individually affected by unital channels.

\textbf{Theorem $\mathbf{4}$.} \textit{ For an $N$-qubit pure state ($N \geq 3)$ between $N - 1$ senders and a single receiver, the sufficient condition for $m$ ($m < N-1$) unital channels acting on different sender subsystems to break dense codeability is $\sum_{i=1}^{m}p_{1}^{(i)} \geq 1$, with $p_1^{(i)}$ being the co-efficient of $\Lambda_1$ in $\Lambda_{UNI}$, applied on $i$-th party.}

\textit{Proof}. Let us first note that for an $N$-qubit pure state, $\rho_N$, involved in a dense coding protocol comprising $N - 1$ senders and a lone receiver, $C(\rho_N) \leq N$, whereas the best classical technique can attain a capacity of $N - 1$. When a unital channel, $\Lambda^{(1)}$, acts on one of the senders (say, the first one, as conveyed by the superscript), the noisy state may be represented as
\begin{eqnarray}
  \nonumber   \rho_{N,\Lambda^{(1)}}&& =  (\Lambda^{(1)} \otimes \mathcal{I}_{2}^{\otimes N - 1})  \rho_N = p^{(1)}_1 \mathbb{I}_2/2 \otimes \rho_{N-1} + \\&&p^{(1)}_2 \rho_N +
 \frac{p^{(1)}_3}{2}(\rho_N + \rho_N^{x_1}) + \frac{p^{(1)}_4}{2}(\rho_N^{x_1} + \rho_N^{y_1}),\nonumber\\
  \label{eq:unital-1}
\end{eqnarray}
where $\rho_{N - i} = \Tr_{1,2,\cdots,i} \rho_N$ is an $(N-i)$-qubit state and for brevity, $\rho_N^{x_i(y_i)} = (\mathbb{I}_2^{\otimes i - 1} \otimes \sigma_x^i (\sigma_y^i) \otimes \mathbb{I}_2^{\otimes N - i}) \rho_N (\mathbb{I}_2^{\otimes i - 1} \otimes \sigma_x^i (\sigma_y^i) \otimes \mathbb{I}_2^{\otimes N - i})$. Using the concavity property of the conditional entropy, one can upper bound the dense coding capacity pertaining to Eq. \eqref{eq:unital-1} as
\begin{eqnarray}
    C(\rho_{N,\Lambda^{(1)}}) \leq p^{(1)}_1 (N - 1) + (1 - p^{(1)}_1) N,
    \label{eq:unital-dcc-1}
\end{eqnarray}
where we have used the fact that states local unitarily connected to $\rho_N$ possess $C(\rho_N) \leq N$, and that $C(\mathbb{I}_2/2 \otimes \rho_{N-1}) \leq N - 1$, together with the normalization condition. In order for $\Lambda^{(1)}$ to reduce the dense coding capacity below the classical threshold, we demand that
\begin{eqnarray}
    \nonumber  p^{(1)}_1 (N - 1) &&+ (1 - p^{(1)}_1) N \leq N - 1  \\
    \implies && p^{(1)}_1 \geq 1\implies p_1^{(1)}=1.
    \label{eq:unital-1-condition}
\end{eqnarray}
Therefore, for a single qubit unital channel acting on an $N$-qubit dense coding resource state, the sufficient condition which dictates the breakdown of its dense codeability is that $\Lambda^{(1)} (.) = \Tr(.)\mathbb{I}_2/2$. In order to facilitate our analysis, we note that the action of $\Lambda^{(1)}$ on $\rho_N$ changes the bound on its dense coding capacity from $N \to p^{(1)}_1 (N - 1) + (1 - p^{(1)}_1) N$, i.e., with probability $p^{(1)}_1$ the capacity bound is reduced by one bit, while with probability $1 - p^{(1)}_1$, it remains unchanged. Following this logic, one can derive the sufficient condition for two unital channels $\Lambda^{(1)} \otimes \Lambda^{(2)}$ to break the dense coding capability of $\rho_N$ as
\begin{eqnarray}
  && \nonumber  C(\rho_{N,\Lambda^{(1)}, \Lambda^{(2)}}) \leq p^{(1)}_1(p^{(2)}_1 (N - 2) + (1 - p^{(2)}_1)\times\\&&(N - 1))
  \nonumber + (1 - p^{(1)}_1)(p^{(2)}_1 (N - 1) + (1 - p^{(2)}_1) N) \\
  && \nonumber = p^{(1)}_1 p^{(2)}_1 (N - 2)  \\
  && \nonumber + (p^{(1)}_1 (1 - p^{(2)}_1) + p^{(2)}_1 (1 - p^{(1)}_1)) (N - 1) \\ 
  && \nonumber + (1 - p^{(1)}_1)(1 - p^{(2)}_1) N \leq N - 1 \\
  && \implies p^{(1)}_1 + p^{(2)}_1 \geq 1.
\end{eqnarray}
%Similarly, upon applying $(\Lambda_1 \otimes \Lambda_2 \otimes \Lambda_3 \otimes \mathbb{I}_2^{\otimes N - 3})$ the dense coding capacity is sufficiently reduced below the classical threshold when the channel parameters satisfy $ p^{(1)}_1 + p^{(2)}_1 + p^{(3)}_1 \geq 1$. 
Inductively, one immediately obtains
\begin{eqnarray}
   && \nonumber  C(\rho_{N,\Lambda^{(1)},\Lambda^{(1)},\cdots,\Lambda^{(m)}})  \leq p^{(1)}_1 p^{(2)}_1 \cdots p^{(m)}_1 (N - m) + \\
   && \nonumber \sum_{k = 1}^{m}(1 - p^{(k)}_1) \prod_{\substack{i = 1 \\ i \neq k}}^{m} p^{(i)}_1 (N - m +1) + \\
   && \nonumber \sum_{k,l = 1}^{m} (1 - p^{(k)}_1)(1 - p^{(l)}_1) \prod_{\substack{i = 1 \\ i \neq k,l}}^{m} p^{(i)}_1 (N - m +2) + \\
   \nonumber && ~~~~~~~~~~~~~~\cdots \cdots \cdots \cdots +  \\
 && \nonumber  (1 - p^{(1)}_1)(1 - p^{(2)}_1) \cdots (1 - p^{(m)}_1) N \leq N - 1 \\
   && \implies p^{(1)}_1 + p^{(2)}_1 + \cdots + p^{(m)}_1 \geq 1.
   \label{eq:DCC-ubital-multi}
    \end{eqnarray}
Therefore, $C(\rho_{N,\Lambda^{(1)},\cdots,\Lambda^{(m)}}) \leq N - 1$, if $\sum_{i=1}^{m}p_{1}^{(i)} \geq 1$. Hence the proof. $\hfill \blacksquare$

\textcolor{black}{\textbf{Note $3$:} For $N=2$, the sufficient condition for a unital qubit channel to be classified as a DBT, in terms of its canonical components, is $p_1\geq\frac{1}{2}$.}

We now shift our attention to those channels that can convert any resource state useless for teleportation while preserving entanglement.

\section{Characterization of Teleportation resource-breaking channels}
\label{sec:TBT}

We consider the {\bf optimal} teleportation protocol in this case, such that $\Lambda \in OTBT$ which trivially implies $\Lambda \in TBT$. We begin with bipartite qudit resource states, and then present conditions to break teleportation in $\mathbb{C}^2 \otimes \mathbb{C}^2$.

\textbf{Theorem $\mathbf{5}$.} \textit{If a group-covariant channel, $\Lambda$, can make any ME state useless for teleportation, it does so for all input pure states.}

%The following conditions are sufficient for a channel, $\Lambda$, to convert all states useless for the teleportation protocol.\\
%(a) It }

\textit{Proof}. As already discussed, an arbitrary pure state $\rho_{\psi}=\ketbra{\psi}{\psi}~\text{on}~\mathbb{C}^d\otimes\mathbb{C}^d$, locally affected by an arbitrary group-covariant channel, $\Lambda$, becomes $(\Lambda\otimes\mathcal{I})\rho_{\psi}=\rho_{\psi,\Lambda}=\Lambda'(\rho_{\phi^+,\Lambda})$ with $\Lambda'\in LOCC$. The singlet fraction of $\rho_{\psi,\Lambda}$ in the optimal teleportation protocol is, therefore,
\begin{eqnarray}
   \nonumber f(\rho_{\psi, \Lambda}) &&= \max_{L \in \text{LOCC}} \bra{\phi^+} L(\Lambda'(\rho_{\phi^+,\Lambda})) \ket{\phi^+} \\
    &&\leq \max_{L\in \text{LOCC}} \bra{\phi^+} L(\rho_{\phi^+,\Lambda}) \ket{\phi^+}.
\end{eqnarray}
 The last inequality follows from $f$ being an LOCC monotone~\cite{Verstraete_PRL_2003}. If $\Lambda$ breaks the teleportation capability of $\rho_{\phi^+}$, $f_{\max}(\rho_{\phi^+,\Lambda}) \leq 1/d\implies f_{\max}(\rho_{\psi,\Lambda}) \leq 1/d~\forall~\ket{\psi}\in\mathbb{C}^d\otimes\mathbb{C}^d$, and therefore, it destroys the quantum advantage of all pure states. $\hfill \blacksquare$

%\textbf{Corollary.} \textit{$\Lambda$ is a {\bf standard} teleportation breaking channel, if it can make any ME state resourceless for the protocol.}

%\textit{Proof.} Any channel, $\Lambda$, is equivalent to the depolarising channel upon twirling operations while retaining the same singlet fraction in the absence of LOCC pre-processing~\cite{Horodecki_PRA_1999}. Since the depolarising channel is group-covariant, Theorem $5$ implies that $\Lambda$ needs to break the {\bf standard} teleportation fidelity of any ME state to be $TBT$. $\hfill \blacksquare$

\textbf{Qubit $TBT$ channels: } The plethora of studies on qubit teleportation~\cite{RajarshiPal_PRA_2014, Horodecki_PRA_1999, Verstraete_PRL_2003} allows us to further investigate $OTBT$ channels in $\mathbb{C}^2 \otimes \mathbb{C}^2$. The following theorem provides a precise characterization of qubit $OTBT$ channels.

\textbf{Theorem 6.} \textit{$\Lambda$ can break the teleportation capability of a  pure two-qubit resource iff it is an entanglement-breaking channel.}

\textit{Proof.}  The optimal singlet fraction for a channel $\Lambda$ is defined as $f_{\max}(\Lambda)= \substack{\max\\{\ket{\psi}\in\mathbb{C}^d\otimes\mathbb{C}^d}}~\substack{\max\\{L\in LOCC}}\braket{\phi^+}{L((\Lambda\otimes\mathcal{I})\rho_{\psi})|\phi^+}$. By definition, $f_{\max}(\Lambda)$ is the maximum over all pure bipartite states affected by $\Lambda$ on one party. Thus $f_{\max}(\Lambda)\leq1/d \implies \Lambda \in OTBT$. In case of a qubit channel, $\Lambda_{qubit}$, $f_{\max}(\Lambda_{qubit}) = 1/2 (1 + \mathcal{N}(\rho_{\phi^+, \Lambda_{qubit}}))$~\cite{RajarshiPal_PRA_2014}, where $\mathcal{N}(.)$ denotes the negativity~\cite{Vidal_PRA_2002}. Therefore, if $f_{\max}(\Lambda_{qubit}) \leq 1/2$, it implies $\mathcal{N}(\rho_{\phi^+, \Lambda_{qubit}}) = 0$, which means $\Lambda_{qubit}$ destroys the entanglement of $\rho_{\phi^+}$. Conversely, if $\Lambda_{qubit}$ is entanglement-breaking, then $\mathcal{N}(\rho_{\phi^+, \Lambda_{qubit}}) = 0 \implies f_{\max}(\Lambda_{qubit}) \leq 1/2$. Thus in $\mathbb{C}^2\otimes\mathbb{C}^2$, $OTBT$ is equivalent to $EBT$. $\hfill \blacksquare$
Recall that all two-qubit entangled states provide quantum advantage in optimal teleportation~\cite{Verstraete_PRL_2003}, which intuitively justifies Theorem $6$. However, it is still an interesting question whether the results also hold in higher dimensions, where such a correspondence between distillable entanglement and teleportation fidelity is not known.
%It has been established that the optimum singlet fraction obtainable from a qubit non-unital channel $\Lambda'$ is given by $f_{\max} = 1/2 (1 + \mathcal{N}(\rho_{\phi^+, \Lambda'}))$, where $\rho_{\phi^+, \Lambda'} = (\Lambda' \otimes \mathbb{I}_2) \rho_{\phi^+}$ and $\mathcal{N}(\sigma)$ denotes the negativity of the qubit state $\sigma$. Therefore, in order that $f_{\max} \leq 1/2$, we must have $\mathcal{N}(\rho_{\phi^+, \Lambda'}) \leq 0$, which further implies that $\Lambda'$ must destroy the entanglement of the maximally entangled state $\rho_{\phi^+}$. From the definition of an entanglement breaking channel, it follows that $\Lambda'$ would also render all other entangled pure states separable. Hence the proof. $\hfill \blacksquare$

Let us now discuss the detection of channels that can preserve resources for quantum tasks.

\section{Construction of witness operators}
\label{sec:witness}

Having established that $DBT(TBT)$ constitutes a convex and compact set, it is possible to identify a channel outside $DBT(TBT)$. By the Hahn-Banach separation theorem, there exists a hyperplane (witness) for every non-$DBT(TBT)$ map that separates them from the set of $DBT(TBT)$ channels. Let $W_1$ and $W_2$ be nonempty, disjoint convex sets in a real Banach space with one of them being compact and the other, closed. There exists a continuous functional $f:W_i\rightarrow \mathbb{R}$ (with $i=1,2$) and $\alpha\in \mathbb{R}$ such that for all pairs $w_1\in W_1$ and $w_2\in W_2$, $f(w_1)<\alpha\leq f(w_2)$. Since a one-element set is compact, we can separate a non-$DBT(\text{non-}TBT)$ map from the set of $DBT(TBT)$ maps by constructing $f$. According to Choi-Jamiolkowski isomorphism~\cite{Choi_LAA_1975,Jamiolkowski_RMP_1972}, we can associate every $(CPTP)$ operation to a specific higher dimensional density operator via a bijective mapping. Notice that each channel in $DBT(TBT)$ corresponds to a specific dual state in the set of non-resourceful states for quantum tasks ($NRSQT$), although there exist states in $NRSQT$ that correspond to dual channels outside the set of $DBT(TBT)$ maps.

Combining these facts, we employ a technique similar to that in Ref.~\cite{Sohail_PLA_2021}, by considering a functional $f_{\{\mathcal{W},\ket{\phi}\}}$ corresponding to $\{\mathcal{W},\ket{\phi}\}$ such that $f_{\{\mathcal{W},\ket{\phi}\}}(\Lambda)=\Tr \left(\mathcal{W}(\Lambda \otimes \mathcal{I})|\phi\rangle\langle\phi|\right)$,
%\begin{equation}
 %   \label{eq:witness_functional}
  %  f_{\{\mathcal{W},\ket{\phi}\}}(\Lambda)=\Tr \left(\mathcal{W}(\Lambda \otimes \mathbb{I})|\phi\rangle\langle\phi|\right),~~~~~~~~
%\end{equation}
where $\Lambda \in\text{CPTP}$, $\ket{\phi}\in S_\mathcal{P}^{\rho_\mathcal{Q}}$, and $\mathcal{W}$ is a witness operator to identify whether a state is resourceful for the aforementioned tasks. It follows that $(i)~ f_{\{\mathcal{W},\ket{\phi}\}}(\Lambda)\geq0~~~\forall~\Lambda\in DBT(TBT)$ and $(ii)~ \text{for}~ \Lambda\notin DBT(TBT)$, $f_{\{\mathcal{W},\ket{\phi}\}}(\Lambda)<0$ for some $\mathcal{W}$ and $\ket{\phi}$. Consequently, $f_{\{\mathcal{W},\ket{\phi}\}}$ can be used as a witness operator to determine whether a given channel is outside $DBT(TBT)$. It is important to note that, we not only need to look for $\mathcal{W}$ but also a suitable $\ket{\phi}$ to witness a channel $\Lambda\notin DBT(TBT)$.

We can, however, provide a more detailed construction of witness operators if the communication protocol (e.g., dense coding or teleportation) or some partial information about the given channel (e.g., possesses group-covariance nature) is specified. For instance, if we consider dense coding protocol, we can utilize $\mathcal{W}=-\log_2(\Lambda\otimes\mathcal{I}|\phi\rangle\langle\phi|)+\mathbb{I}\otimes\log_2(\Tr_\mathcal{S}(\Lambda\otimes\mathcal{I}|\phi\rangle\langle\phi|))$~\cite{Vempati_PRA_2021} and only need to search for $\ket{\phi}\in\mathbb{C}^{d_\mathcal{S}}\otimes\mathbb{C}^{d_\mathcal{R}}$ to determine whether a given channel $\Lambda \notin DBT$. Further, for group-covariant channels, $\ket{\phi} = \ket{\phi^+}$. If we consider the optimal teleportation scheme involving two-qubit resource states, $\ket{\phi}$ can be safely taken as $\ket{\phi^+}$ with $\mathcal{W}$ being the optimal witness operator for two-qubit NPT entangled states~\cite{Terhal_TCS_2002}. We discuss examples of witnesses for the depolarizing channel in Ref.~\cite{Supp}.

\section{Conclusion}
\label{sec:conclu}

With the advent of the second quantum revolution, shared entanglement has emerged as one of the pivotal resources for quantum protocols. Although it is well known that entanglement is necessary for providing quantum advantage in several tasks, it is not sufficient. For a fixed quantum process, we developed a framework characterizing the corresponding resource-breaking channels ($PBT$), i.e., any resource state for a specific scheme becomes useless after the application of the quantum channel. In particular, we analyzed quantum dense coding (DC) and teleportation (QT), and investigated the \textit{non-trivial} sets of teleportation ($TBT$) and dense coding  ($DBT$) resource-breaking channels which are, in general, different from entanglement-breaking channels ($EBT$).
%(\textit{DBT}), where \textit{non trivial} in the sense that they are different from entanglement-breaking channels (\textit{EBT}).

We have proved that both the sets, $TBT$ and $DBT$, are convex and compact and identified a certain class of extreme classical-quantum channels as their extreme points. Further, we presented sufficient conditions for a channel to be $TBT$ or $DBT$. Group-covariant channels with the ability to transform a maximally entangled state into a resourceless one, are either $OTBT$ or $DBT$. In the case of standard dense coding, we provided a sufficient condition on unital qubit channels, that break the dense codeability of any resource state. In contrast, we demonstrated that the set $OTBT$ for qubits is equivalent to that of qubit $EBT$ channels. However, in the case of higher dimensions, such equivalence is still an open question. We have also illustrated how to construct witness operators capable of identifying standard non-$TBT$(non-$DBT$) maps.

Our work has a consequential impact on several protocols, viz., {\it conditional state transfer}~\cite{Fedorov_PRL_2023}, {\it optimal resource-efficient use of quantum channels}~\cite{Adesso_PRL_2018}, {\it benchmarking quantum supremacy}~\cite{Cavalcanti_PRL_2017}, {\it resource distillation}~\cite{Takagi_PRL_2024}, {\it many-body teleportation}~\cite{Agarwal_PRL_2023}, {\it delocalized interactions}~\cite{Paige_PRL_2020}, {\it multiplexed quantum communication}~\cite{Shi_PRL_2020, Chen_PRL_2021}, and the {\it study of natural mode entanglement in ultracold gases}~\cite{Libby_PRL_2009} which rely on dense coding and teleportation as their fundamental ingredient. The concept of $PBT$ has also been recently applied to quantum computation~\cite{Patra_arXiv_2024}. It also highlights an important property of group-covariant channels. Furthermore, since both negative conditional entropy and the singlet fraction are key resources for many tasks~\cite{Horodecki_ComMathPhys_2007, Horodecki_Nat_2005, Devetak_RoyalSociety_2005, Bennett_PRL_1996, Zhao_JPhysA_2010, Zhao_PRA_2015, Grondalski_PhysLettA_2002, Patro_EurPhysJD_2022, Nirman_PRL_2011}, our work identifies channels that destroy such resources, paving the way for a dynamical resource theory of preservability~\cite{Hsieh_Quantum_2020}, wherein such channels are considered to be free objects.

Our study further raises several interesting questions such as, whether the set of $PBT$ maps concerning other quantum protocols also form a convex and compact set, and what particular properties they possess or whether there exists any hierarchy among them. Moreover, it will be fascinating to find the existence of a non-maximally entangled state for non-group-covariant channels, upon which their action can determine whether it belongs to $TBT$($DBT$). Last, but not least, channels that can break the capacity of states under generic pre-processing are also yet to be identified for the optimal dense coding scheme. The structure and the analysis carried out in this work, therefore, open up the possibility for further examinations in this direction.

\begin{acknowledgments} 
We thank Alessandro Ferraro for discussions during QIPA 2023. We acknowledge the support from the Interdisciplinary Cyber-Physical Systems (ICPS) program of the Department of Science and Technology (DST), India, Grant No.: DST/ICPS/QuST/Theme- 1/2019/23.  We acknowledge the use of \href{https://github.com/titaschanda/QIClib}{QIClib} -- a modern C++ library for general-purpose quantum information processing and quantum computing (\url{https://titaschanda.github.io/QIClib}).  This research was supported in part by the 'INFOSYS scholarship for senior students' R.G. acknowledges funding from the HORIZON-EIC-$2022$-PATHFINDERCHALLENGES-$01$ program under Grant Agreement No.~$10111489$ (Veriqub). Views and opinions expressed are however those of the authors only and do not necessarily reflect those of the European Union. Neither the European Union nor the granting authority can be held responsible for them..
\end{acknowledgments}

\begin{figure*}[h!]
    \includegraphics[width=\linewidth]{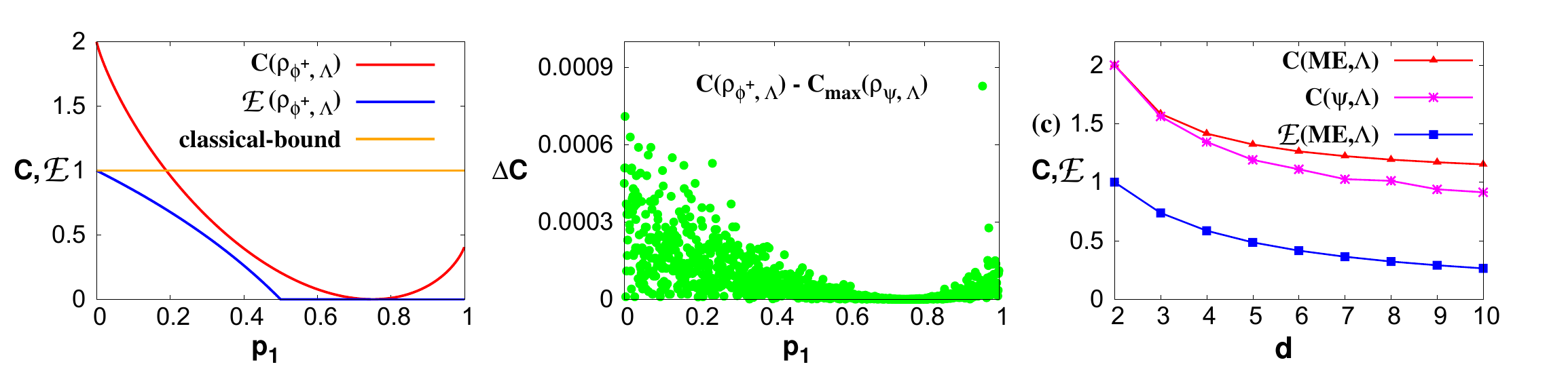}
\captionsetup{justification=Justified,singlelinecheck=false}
     \caption{\textbf{Group-covariant dense coding resource-breaking channel in $\mathbb{C}^2 \otimes \mathbb{C}^2$ }. (a). The dense coding capacity $C$ (ordinate) against the channel parameter $p_1$(abscissa). The red (dark) circles represent the capacity for $\rho_{\phi^+, \Lambda}$ whereas the orange (light) line represents the classical bound, $1$ bit. It is observed that the channel can reduce the capacity of the state $\ket{\phi^+}$ below the classical threshold. (b). $\Delta C$ (ordinate) against the channel parameter $p_1$ (abscissa) in green (light) circles. It is apparent that the difference is always positive, thereby highlighting that $\Lambda \in DBT$ if $C(\rho_{\phi^+, \Lambda}) \leq 1$. (c). The entanglement $E$ (ordinate) of $\rho_{\phi^+, \Lambda}$, quantified by the logarithmic negativity \cite{Plenio_PRL_2005}, is shown against the channel parameter $p_1$ (abscissa). It is evident that $\Lambda \notin EBT$ when $p_1 > 0.5$. Quantities are computed by simulating $10^5$ random pure sates, $\rho_{\psi,\Lambda}$, on which the channel, $\Lambda$, acts. The x-axis is dimensionless whereas the y-axis is in bits for (a). and (b). and in e-bits for (c). }
    \label{fig:gc-dbt}
    \end{figure*}

\appendix

\section{Notations and Definitions}
\label{app:pre}
We here provide a comprehensive overview of the notations that are used throughout the manuscript. In addition, we also give definitions and properties of certain exemplary channels which are useful for the discussions. For completeness, we heuristically discuss the protocols of dense coding and quantum teleportation towards the end of this part.
\subsubsection{Mathematical preliminaries}
Throughout this paper, we denote $d$-dimensional complex Hilbert spaces by $\mathbb{C}^d$ while the real space is defined as $\mathbb{R}$. The tensor product of two complex spaces, of dimension $d_1$ and $d_2$ respectively, is denoted as $\mathbb{C}^{d_1} \otimes \mathbb{C}^{d_2}$. Let the computational basis vectors in $\mathbb{C}^d$ be $\ket{i}$ or $\ket{j}$, with $i,j \in \{0,1,\cdots,d-1\}$. Pure states are represented using smaller case Greek letters such as $\ket{\phi}, \ket{\psi}$ whereas the corresponding density matrices, which are Hermitian, positive semidefinite bounded linear operators with unit trace, are denoted as $\rho_{\phi},$ and $\rho_{\psi}$ respectively. Arbitrary density matrices are described without the subscript, e.g., $\chi$.
The maximally entangled (ME) state in $\mathbb{C}^d \otimes \mathbb{C}^{d}$ takes the form $\ket{\phi^+} = 1/\sqrt{d} \sum_{i = 0}^{d - 1} \ket{i} \ket{i}$. In the two-qubit Hilbert space, $d = 2$, $\ket{\phi^+}$ represents one of the four Bell states, while the other three are $\ket{\phi^-} = 1/\sqrt{2} (\ket{00} - \ket{11}), \ket{\psi^\pm} = 1/\sqrt{2} (\ket{10} \pm \ket{01})$. Density matrices corresponding to other ME states are represented by $\rho_{\text{ME}}$. Moreover, states that are not resourceful for information processing tasks belong to the set of ``\textit{non-resourceful states for quantum tasks}'', $NRSQT$.

We depict unitary operators with the capital letters $U, V$, while $M$ portrays generalized measurements or positive operator valued measurements (POVMs). Note that $U^\dagger$ represents the Hermitian-conjugate of $U$. The $d$-dimensional identity matrix is denoted as $\mathbb{I}_d$ and $\sigma_x, \sigma_y ~ \text{and} ~ \sigma_z$ are the usual Pauli matrices. A local operation and classical communication $(LOCC)$ operator is represented by $L$.

Quantum channels, which are completely positive trace-preserving ($CPTP$) maps, are described by $\Lambda$. Note that a channel is said to be unital if it maps the identity operator to itself, i.e., $\Lambda_{UNI} \in CPTP$ is unital $\implies \Lambda_{UNI}(\mathbb{I}_d) = \mathbb{I}_d$. Channels that do not follow this relation are known as non-unital channels. The action of a channel on a subsystem of a composite state, $\rho_{\phi}$ acting on $\mathbb{C}^d \otimes \mathbb{C}^d$, can be described as $\rho_{\phi, \Lambda} = (\Lambda \otimes \mathcal{I}_d) \rho_{\phi}$. 

The von-Neumann entropy of a quantum state, $\rho$, is defined as $S(\rho) = -\Tr(\rho \log_2 \rho)$ \cite{nielsen_2010}. For a composite system, $\rho_{AB}$ acting on $\mathbb{C}^d \otimes \mathbb{C}^d$ shared between two parties $A$ and $B$, the conditional von-Neumann entropy reads as $S_{A|B}(\rho_{AB}) = S(\rho_{AB}) - S(\rho_B)$. Furthermore, the relative entropy between two states $\rho$ and $\chi$ is given by $S(\rho || \chi) = \Tr(\rho (\log_2 \rho - \log_2 \chi))$. Note that the relative entropy satisfies the property of monotonicity under $CPTP$ maps, i.e., $S(\rho_{\Lambda} || \chi_{\Lambda}) \leq S(\rho || \chi)$ \cite{Lindblad_Springer_1975}, and is jointly convex with respect to its arguments, i.e., $S(\sum_i p_i \rho_i || \sum_i p_i \chi_i) \leq \sum_i p_i S(\rho_i || \chi_i)$ \cite{nielsen_2010}.

\subsubsection{Dense coding and Teleportation protocols}
We lay down the framework of two exemplary quantum communication protocols, viz, dense coding (DC) and teleportation (QT), along with their figures of merit.

\begin{enumerate}
    \item \textbf{Dense Coding.} The dense coding protocol \cite{Bennett_PRL_1992} involves sending classical information from $N -1$ senders, $\mathcal{S}_1, \cdots, \mathcal{S}_{N - 1}$, of dimension $d_{\mathcal{S}_1}, \cdots, d_{\mathcal{S}_{N - 1}}$ to a receiver $\mathcal{R}$, of dimension $d_{\mathcal{R}}$, where an $N$-party resource state, $\rho_{\mathcal{S}_1 \cdots \mathcal{S}_{N - 1} \mathcal{R}}$ is shared among them. Upon encoding through orthogonal unitary operators, with optimal probabilities \cite{Hiroshima_JPA_2001}, each sender sends the corresponding quantum system to the receiver through a noiseless quantum channel. The receiver performs an optimal measurement to gather maximum classical information encoded by the senders, whereafter the dense coding capacity ($DCC$) of a given state, $\rho_{\mathcal{S}_1 \cdots \mathcal{S}_{N - 1} \mathcal{R}}$ is given by \cite{Bruss_PRL_2004}
    \begin{eqnarray}
    && C = \max\Big[\log_2d_{\mathcal{S}_1} \cdots d_{\mathcal{S}_{N - 1}}, \nonumber\\ 
      &&  \log_2d_{\mathcal{S}_1} \cdots d_{\mathcal{S}_{N - 1}} + S(\rho_{\mathcal{R}}) - S(\rho_{\mathcal{S}_1 \cdots \mathcal{S}_{N - 1} \mathcal{R}})\Big],
      \label{eq:dcc}
    \end{eqnarray}
    where $\rho_{\mathcal{R}} = \Tr_{\mathcal{S}_1, \cdots, {\mathcal{S}_{N - 1}}} \rho_{\mathcal{S}_1 \cdots \mathcal{S}_{N - 1} \mathcal{R}}$ is the receiver's subsystem. 
    %In Eq. \eqref{eq:dcc:rel-ent}, $\rho^j$ denotes the state after the message $j$ is encoded with probability $q_j$ using orthogonal unitaries $U^{j}_{\mathcal{S}_i}$ by each sender $\mathcal{S}_i$ and $\overline{\rho} = \sum_j q_j \rho^j$. In the case of optimal unitaries and probabilities, Eq. \eqref{eq:dcc:rel-ent} reduces to Eq. \eqref{eq:dcc}.

    It may be the case that the channel through which the encoded subsystems are sent is noisy. In this situation, the dense coding capacity takes the form as \cite{Shadman_NJP_2010}
    \begin{eqnarray}
        &&  \nonumber   \tilde{C} = \max\Big[\log_2d_{\mathcal{S}_1} \cdots d_{\mathcal{S}_{N - 1}}, \\ 
      && \nonumber  \log_2d_{\mathcal{S}_1} \cdots d_{\mathcal{S}_{N - 1}} + S(\rho_{\mathcal{R}}) - S((\tilde{\Lambda}_{\mathcal{S}_1 \cdots \mathcal{S}_{N-1}} \otimes \mathcal{I}_{\mathcal{R}}) \times\\
      && \nonumber (U_{\mathcal{S}_1} \otimes \cdots \otimes U_{\mathcal{S}_{N-1}}\otimes \mathbb{I}_{d_{\mathcal{R}}}\rho_{\mathcal{S}_1 \cdots \mathcal{S}_{N - 1} \mathcal{R}} \times \\
      &&  U^\dagger_{\mathcal{S}_{1}} \otimes \cdots \otimes U^\dagger_{\mathcal{S}_{N-1}} \otimes \mathbb{I}_{d_{\mathcal{R}}}))\Big],
      \label{eq:noisy_dcc}
    \end{eqnarray}
    with $U_{\mathcal{S}_i}$ being unitaries applied by the sender $\mathcal{S}_i$ to minimize the entropy of the state upon the action of the noisy channel $\tilde{\Lambda}$.

    Note that the capacity attainable via classical means alone is  $\log_2d_{\mathcal{S}_1} \cdots d_{\mathcal{S}_{N - 1}}$ bits, which implies that the quantum advantage can be achieved if and only if $S_{\mathcal{S}|\mathcal{R}} \equiv S_{\mathcal{S}_1 \cdots \mathcal{S}_{N - 1}| \mathcal{R}} = S(\rho_{\mathcal{R}}) - S(\rho_{\mathcal{S}_1 \cdots \mathcal{S}_{N - 1} \mathcal{R}}) > 0$ (or $S(\rho_{\mathcal{R}}) - S((\tilde{\Lambda} \otimes \mathcal{I}) (U_{\mathcal{S}_1} \cdots U_{\mathcal{S}_{N-1}}\rho_{\mathcal{S}_1 \cdots \mathcal{S}_{N - 1} \mathcal{R}}U^\dagger_{\mathcal{S}_{1}} \cdots U^\dagger_{\mathcal{S}_{N-1}})) >0$  in the noisy case). 
    %We refer to channels $\Lambda (\tilde{\Lambda})$ as dense coding resource-breaking channels when the quantum advantage vanishes for the state $\rho_{\mathcal{S}_1 \cdots \mathcal{S}_{N - 1} \mathcal{R}, \Lambda (\tilde{\Lambda})}$. 

    \item \textbf{Teleportation.} Quantum teleportation \cite{Bennett_PRL_1993} refers to the transfer of an unknown state, say $\rho_{\phi}$ acting on $\mathbb{C}^d$, from one location to another with the help of a resource state, say $\rho_{\mathcal{S} \mathcal{R}}$ acting on $\mathbb{C}^d \otimes \mathbb{C}^d$ and a finite amount of classical communication. The figure of merit for the protocol is given by the fidelity, $F$, between $\rho_{\phi}$ and the output state. In the case of a qubit-teleportation, the optimal output fidelity (i.e., for the optimal teleportation protocol) is bounded as \cite{Horodecki_PLA_1996}
    \begin{equation}
        F_{\max} \leq \frac{1}{2} (1 + \frac{1}{3} \Tr \sqrt{C^\dagger C}),
        \label{eq:2-fid}
    \end{equation}
    where the elements of the matrix $C$ represent classical correlators, given by $C_{ij} = \Tr(\rho_{\mathcal{S} \mathcal{R}} \sigma_i \otimes \sigma_j)$ with $\{i,j\} = \{x, y, z\}$. In arbitrary dimensions, the fidelity can be expressed in terms of the singlet-fraction $f$ as \cite{Horodecki_PRA_1999}
    \begin{eqnarray}
    && F_{\max} = \frac{ d~f+ 1}{d + 1} ~~ \text{with} \\
       && f_{\max} = \max \Big[ \frac{1}{d},~ \max_{L} \bra{\phi^+} L(\rho_{\mathcal{S} \mathcal{R}}) \ket{\phi^+} \Big],
    \end{eqnarray}
    where the maximization is performed over LOCC, $L$. Note that, in case of {\bf {\bf standard}} teleportation protocol, LOCC should be replaced by local unitary map. The classical threshold in the teleportation process is $2/(d+1)$ \cite{Massar_PRL_1995}, and the quantum advantage is obtained if and only if $F_{\max} > 2/(d+1)$, which is possible only when $f_{\max} \geq 1/d$. 
    %Therefore, a channel $\Lambda$ is said to be a teleportation resource-breaking channel when $f_{\max}(\rho_{\mathcal{S} \mathcal{R},\Lambda}) \leq 1/d~~\forall~~\rho_{\mathcal{S} \mathcal{R}}$ acting on $\mathbb{C}^d \otimes \mathbb{C}^d$.
\end{enumerate}

\subsubsection{Paradigmatic relevant channels}
The definitions and properties of certain quantum channels, useful for analyzing $PBT$ maps in the context of $DC$ and $QT$, are elucidated in this subsection.

\begin{itemize}
\item $\mathbf{\Lambda_{NCVE}}$ \cite{Vempati_Quantum_2022} :- A channel $\Lambda:\mathbb{C}^{A} \otimes \mathbb{C}^{B} \to \mathbb{C}^{C} \otimes \mathbb{C}^{D}$ is said to be conditional von-Neumann entropy non-decreasing ($NCVE$) if for all $\rho_{AB}$ acting on $\mathbb{C}^{A} \otimes \mathbb{C}^{B}$ and for all $\rho_{CD} = \Lambda (\rho_{AB})$ acting on $\mathbb{C}^{C} \otimes \mathbb{C}^{D}$, $S_{A|B}(\rho_{AB}) \leq S_{C|D}(\rho_{CD})$. When the input and output spaces of such a channel are the same, they are denoted as $NCVE(A|B)$.

\item $\mathbf{\Lambda_{UNI(A|B)}}$ \cite{Vempati_Quantum_2022} :-A channel $\Lambda:\mathbb{C}^{A} \otimes \mathbb{C}^{B} \to \mathbb{C}^{A} \otimes \mathbb{C}^{B}$ is said to be A-unital ($UNI(A|B)$) if for all $\rho_B$ acting on $\mathbb{C}^B,$ $\Lambda (\frac{\mathbb{I}_A}{d_A}\otimes\rho_{B})=\frac{\mathbb{I}_A}{d_A}\otimes \rho'_B$ for some $\rho'_B$ acting on $\mathbb{C}^B$. Note that $\Lambda \in UNI(A|B)$ $\Longleftrightarrow$ $\Lambda \in NCVE(A|B)$.

\item $\mathbf{\Lambda_{\text{EBT}}}$ \cite{Horodecki_RMP_2003}:-  Entanglement breaking channels ($EBT$), which map any entangled state to separable ones, have the Holevo form $\Lambda_{\text{EBT}}(\rho) = \sum_k R_k \Tr F_k \rho$ (also known as {\it measure-and-prepare channel}) \cite{Holevo_RMS_1998}.  The operational interpretation of this channel can be described as one party performing a measurement $\{F_k\}_k$ with the effect operators $0\leq F_k\leq\mathbb{I}_d$, such that $\sum_dF_k=\mathbb{I}_d$, on its input state $\rho$, sending the outcome $k$ to the other through a classical channel, who subsequently prepares a state $R_k$. If $R_k = \ket{k} \bra{k}$, it is known as a \textit{quantum-classical} channel. Conversely, {\it classical-quantum} ($CQ$) channels are  given as $\Lambda_{\text{CQ}}(\rho)=\sum_k R_k \langle k|\rho|k\rangle$, i.e., $F_k = \ket{k}\bra{k}$ in $\Lambda_{\text{EBT}}$. If $R_k=\ket{\psi_k}\bra{\psi_k}$ is a pure state, $\Lambda_{\text{CQ}}$ will be called an extreme $CQ$ channel. In Ref. \cite{Horodecki_RMP_2003}, it was shown that these extreme $CQ$ maps are also extreme $EBT$ maps. Besides this, if $\langle\psi_i|\psi_j\rangle \neq 0~\forall \{i,j\}$, they are the extreme points of the set of $CPTP$ maps.

\item $\mathbf{\Lambda_{\text{GC}}}$ \cite{Memarzadeh_PRR_2022}:- For any compact (or finite) group $G$, and for every element $g \in G$, let $U(g)$ and $V(g)$ be two $d$-dimensional unitary representations of $G$ on $\mathbb{C}^d$. A channel, $\Lambda$, is said to be group-covariant ($GC$) with respect to these representations if for any $\rho$ acting on $\mathbb{C}^d$ and for all $g \in G$,
\begin{eqnarray}
    \Lambda(U(g) \rho U(g)^\dagger) = V(g) \Lambda(\rho) V(g)^\dagger.
    \label{eq:grp-cov}
\end{eqnarray}
The channel $\Lambda$ is said to be irreducibly covariant if $U(g)$ and $V(g)$ are irreducible representations of $g \in G$ \cite{Nuwairan_IJM_2014, Mozrzymas_JMP_2017, Chang_RMP_2022}.
\end{itemize}

{\it Example of group-covariant channels.} Consider $U$ and $V$ to be two unitary representations of a compact (or finite) group $G$. In general, the unitary representations $U(g)$ and $V(g)$ for $g\in G$ may have different dimensions, say $d_U$ and $d_V$ respectively. In case the group under consideration is $SU(2)$, a channel $\Lambda$ is said to be $SU(2)_{(d_U,d_V)}$  covariant, if, $\forall~ g \in SU(2)$
\begin{equation}
    \Lambda(U(g) \rho U(g)^\dagger) = V(g) \Lambda(\rho) V(g)^\dagger,
\end{equation}
for any $\rho ~\text{acting on} ~ \mathbb{C}^{d_U}$ and $\rho_\Lambda ~\text{acting on} ~ \mathbb{C}^{d_V}$, i.e., the channel $\Lambda : \mathbb{C}^{d_U} \to \mathbb{C}^{d_V}$ \cite{Nuwairan_IJM_2014}. It has been shown that the collection of ``$SU(2)$\textit{-Clebsch-Gordan}'' channels is a basis of all $SU(2)_{(d_U,d_V)}$ covariant channels. Therefore, one can construct an indefinite number of such group-covariant channels through $CPTP$ linear combinations of the $SU(2)$-Clebsch-Godran channels. Specifically, when $d_U = d_V = 3$, the set of  $SU(2)_{(3,3)}$ covariant channels is given by
\begin{equation}
    \Lambda_{GC-3,3} = (1 - p - p') \Lambda_0 + p \Lambda_2 + p' \Lambda_4,
\end{equation}
where $\Lambda_0, \Lambda_2, \Lambda_4 \in \{SU(2)-\text{Clebsh-Gordan}\}$, and $0 \leq  p + p' \leq 1$. The corresponding Kraus operators are specified in Ref. \cite{Chang_RMP_2022}. Remarkably, the aforementioned set also contains the unitary conjugates of the well-known ``\textit{Werner-Holevo}'' channels \cite{Werner_JMP_2002}, and the space of $\Lambda_{GC-3,3}$ is evidently spanned by $\mathbb{I}_3$ and unitary conjugates of extremal Werner-Holevo channels in $\mathbb{C}^3$. As a consequence, the channels relevant to our discussion can be constructed through their convex combination. It is worthwhile to mention here that the Werner-Holevo channels in $\mathbb{C}^d$ themselves, are unital and covariant with respect to $SU(d)$ \cite{Wolf_NJP_2005}. Note that the $d$-dimensional depolarising channel is another unital group-covariant channel, being covariant with respect to $d \times d$ unitary representations of $SU(2)$.

%\section{Proof of Theorem $\mathbf{1}$: Properties of dense coding and teleportation resource-breaking channels}
%\label{sec:properties_of_BT}
%Let us describe the ubiquitous features of the {\bf {\bf standard}} dense coding and teleportation resource-breaking channels, independent of the dimension in which they act. All the properties presented here hold for both $DBT$ and $TBT$, and we will collectively refer to these channels as $PBT_{D,T}$.

\section{Example of a group-covariant $DBT$ channel}
\label{app:gc-dcc-break}

A channel covariant with respect to the $3$-dimensional unitary representation of $SU(2)$ can be written as
\begin{equation}
    \Lambda = (1 - p_1 - p_2) \Phi_0 + p_1 \Phi_2 + p_2 \Phi_4,
\end{equation}
where $\Phi_0, \Phi_2$, and $\Phi_4$ are $SU(2)$-Clebsch-Gordan channels. The corresponding Kraus operators are specified in Ref. \cite{Chang_RMP_2022}. We randomly simulate values of $p_1$ and $p_2$ in $[0,1]$ from a uniform distribution subject to $p_1 + p_2 \leq 1$. This gives us a set of group-covariant channels in $\mathbb{C}^3$ for each value of $\{p_1,p_2\}$. We apply each such channel to the ME state $\ket{\phi^+} \in \mathbb{C}^3 \otimes \mathbb{C}^3$ and also on $10^5$ Haar-uniformly generated pure two-qutrit states. The dense coding capacity of $\rho_{\phi^+, \Lambda}$ is plotted with respect to the channel parameter in Fig. \ref{fig:gc-dbt}(a). We observe that the channel can break the dense coding capacity of the ME state. The difference between the $DCC$ for the noisy ME state and the maximum capacity out of the random states, $\Delta C$, is depicted with respect to the channel parameter $p_1$ in Fig. \ref{fig:gc-dbt}(b). We observe that the difference is always positive, i.e., the capacity of any pure state passed through the channel is always less than that of $\rho_{\phi^+, \Lambda}$. Thus the channel can break the {\bf {\bf standard}} dense codeability of all pure states if it can do so for $\ket{\phi^+}$. Furthermore, it has been established that the channel is $EBT$ if $p_1 \leq 1/2$ and $p_1 + p_2 \geq 2/3$. Our figure illustrates that $\Lambda$ can be $DBT$ in nature even in regions where it cannot break the entanglement of its input states (see Fig. \ref{fig:gc-dbt} (c) where $\mathcal{E}(\rho_{\phi^+, \Lambda}) \neq 0$ for $p_1 > 0.5$ and Fig. \ref{fig:gc-dbt} (a) where the dense coding capacity of $\rho_{\phi^+, \Lambda} \leq \log_2 3$ for $p_1 > 0.5$). This is an exemplary demonstration of one of the principal points of study, that there exist channels that can break quantum information processing protocols without eliminating the resource that is necessary for obtaining quantum advantage.

\bibliographystyle{apsrev4-1}
\bibliography{DCB2}

\end{document}